\documentclass[11pt,sort&compress]{article}
\usepackage[a4paper, left=2cm,right=2cm,top=2cm,bottom=2cm]{geometry}
\usepackage[T1]{fontenc}
\usepackage{mlmodern}
\usepackage{bm}
\usepackage{authblk}
\usepackage{amsthm}
\usepackage{amssymb}
\usepackage{mathrsfs}
\usepackage{amsfonts}
\usepackage{amsmath,empheq,fixmath}
\usepackage{nicefrac,xfrac}
\usepackage{multirow}
\usepackage{array}
\usepackage{booktabs}
\usepackage{csvsimple}
\usepackage[dvipsnames]{xcolor}
\usepackage{graphicx}
\usepackage{subfiles}
\usepackage{lipsum}
\usepackage[justification=centering]{caption}
\usepackage{subcaption}
\usepackage{accents}
\usepackage{enumerate}
\usepackage{enumitem}
\usepackage{hyperref}
\usepackage{cleveref}
\hypersetup{
	colorlinks=true,
	linkcolor={blue},
	citecolor={blue},
	urlcolor={blue},
	breaklinks=true,
	plainpages=true
}
\usepackage{calc}
\usepackage{siunitx}
\usepackage{cite}
\usepackage{tikz} 
\usetikzlibrary{backgrounds,shapes,arrows,arrows.meta,positioning,intersections,quotes}
\usepackage{tcolorbox}
\usepackage{pgfplots}
\pgfplotsset{compat=newest,scaled y ticks=true} 
\usepgfplotslibrary{units} 
\usepackage{listings}
\definecolor{codegreen}{rgb}{0,0.6,0}
\definecolor{codegray}{rgb}{0.5,0.5,0.5}
\definecolor{codepurple}{rgb}{0.58,0,0.82}
\definecolor{backcolour}{rgb}{0.95,0.95,0.92}
\lstdefinestyle{mystyle}{
	backgroundcolor=\color{backcolour}, 
	commentstyle=\color{codegreen},
	keywordstyle=\color{magenta},
	stringstyle=\color{codepurple},
	basicstyle=\ttfamily\footnotesize,
	breakatwhitespace=false, 
	breaklines=true, 
	captionpos=b,
	keepspaces=true,
	showspaces=false,
	showstringspaces=false,
	showtabs=false,
	tabsize=1
}
\newlist{myitemize}{itemize}{1}
\setlist[myitemize]{
	label={\textbullet},
	leftmargin=4em,
	itemsep=0.25em,
	topsep=0.25em,
	parsep=0pt
}
\allowdisplaybreaks
\providecommand{\tikzexternaldisable}{}
\providecommand{\tikzexternalenable}{}

\theoremstyle{definition}

\newtheorem{theorem}{\textbf{Theorem}}[section]

\newtheorem{proposition}{\textbf{Proposition}}
\newtheorem{remark}[theorem]{\textbf{Remark}}

\newcommand{\meisu}{Department of Mechanical Engineering, Iowa State University, Ames, IA 50011, USA}
\newcommand{\tracisu}{Translational AI Center, Iowa State University, Ames, IA 50011, USA}

\newcommand{\TUM}{Technical University of Munich, School of Engineering and Design, Professorship for Computational Fluid Dynamics, Boltzmannstr. 15, Garching, 85748, Germany}

\newcommand{\spDom}{\Omega} 

\newcommand{\xcomp}[1]{x_{#1}}
\newcommand{\xii}{\xcomp{i}}
\newcommand{\ucomp}[1]{u_{#1}}

\newcommand{\ui}{\ucomp{i}}

\newcommand{\xvec}{\mvec{x}}

\newcommand{\uf}{{\uvec}_f}

\newcommand{\pf}{p_f}
\newcommand{\uprime}{\uvec'}
\newcommand{\pprime}{p'}
\newcommand{\avec}{\mvec{a}}
\newcommand{\bvec}{\mvec{b}}
\newcommand{\fvec}{\mvec{f}}

\newcommand{\hvec}{\mvec{h}}
\newcommand{\uvec}{\mvec{u}}

\newcommand{\vvec}{\mvec{v}}
\newcommand{\wvec}{\mvec{w}}

\newcommand{\avech}{\mvec{a}_h}
\newcommand{\uvech}{\mvec{u}_h}
\newcommand{\vvech}{\mvec{v}_h}

\newcommand{\zerovec}{\bm{0}}
\newcommand{\ph}{p_h}
\newcommand{\qh}{q_h}

\newcommand{\resMom}{\mvec{R}_m}
\newcommand{\resCon}{\mvec{R}_c}

\newcommand{\partialder}[2]{\frac{\partial #1}{\partial #2}}

\newcommand{\timederShort}{\partial_t}

\newcommand{\taum}{\tau_m}
\newcommand{\tauc}{\tau_c}

\newcommand{\visco}{\nu}
\newcommand{\pder}[1]{\partialder{p}{\xii}}
\newcommand{\normalvec}{\hat{\mvec{n}}}

\newcommand{\uvechN}{\mvec{u}_h^n}

\newcommand{\phN}{\ph^n}
\newcommand{\fvecN}{\mvec{f}^n}
\newcommand{\gvecN}{\mvec{g}^n}

\newcommand{\vh}{v^h}

\newcommand{\spaceV}{\mvec{V}}
\newcommand{\spaceVd}{\mvec{V}^D}
\newcommand{\spaceVh}{{\spaceV}_h}
\newcommand{\spaceVhd}{\spaceVh^D}
\newcommand{\spaceQ}{Q}
\newcommand{\spaceQh}{{\spaceQ}_h}
\newcommand{\spaceTcV}{\mathcal{V}}
\newcommand{\spaceTcVd}{\mathcal{V}^D}
\newcommand{\spaceTcVh}{{\spaceTcV}_h}
\newcommand{\spaceTcVhd}{\spaceTcVh^D}
\newcommand{\spaceTcQ}{\mathcal{Q}}
\newcommand{\spaceTcQh}{{\spaceTcQ}_h}

\newcommand{\mesh}{K^h}

\newcommand{\grad}{{\nabla}}
\newcommand{\divergence}{\grad\cdot}
\newcommand{\ddiv}[1]{#1\cdot\grad}

\newcommand{\adiv}{\avec\cdot\grad}

\newcommand{\divGreek}{\theta}

\newcommand{\laplacian}{\Delta}

\newcommand{\mvec}[1]{{\bm{#1}}}
\newcommand{\mmat}[1]{{\bm{#1}}}

\newcommand{\intSpace}{\int_{\spDom}}

\newcommand{\dx}{{d}x}
\newcommand{\dxv}{{d}\xvec}
\newcommand{\ds}{{d}S}

\newcommand{\inner}[2]{\left( #1, #2 \right)}
\newcommand{\innerB}[2]{\left( #1, #2 \right)_{\partial \Omega}}

\newcommand{\norm}[1]{\left\| #1 \right\|}

\newcommand{\vertiii}[1]{{\left\vert\kern-0.25ex\left\vert\kern-0.25ex\left\vert #1 \right\vert\kern-0.25ex\right\vert\kern-0.25ex\right\vert}}

\newcommand{\oneOver}[1]{\frac{1}{#1}}
\newcommand{\half}{\frac{1}{2}}
\newcommand{\halfnice}{\nicefrac{1}{2}}

\newcommand{\bdfop}{\mathcal{D}_{\dt}}

\newcommand{\opCAdv}[1][\avec]{C_{#1}}

\newcommand{\opDAdv}[1][\avec]{D_{#1}}

\newcommand{\opSAdv}[1][\avec]{S_{#1}}

\newcommand{\opConvForm}[2][\avec]{(\ddiv{#1})#2}
\newcommand{\opDivForm}[2][\avec]{\divergence(#2{#1}^T)}

\newcommand{\opSkewForm}[2][\avec]{(\ddiv{#1})#2 + \halfnice(\divergence {#1})#2}

\newcommand{\opAdvGen}[2][\avec]{\mathcal{M}_{#1,#2}}
\newcommand{\opAdvGenStar}[2][\avec]{\mathcal{M}_{#1,#2}^{*}}
\newcommand{\dvc}{s}

\newcommand{\dt}{\Delta t}

\newcommand{\cinv}{C_{I}}

\newcommand{\colref}[2]{\hyperref[#2]{#1~\ref*{#2}}}
\newcommand{\secref}[1]{\colref{Section}{#1}}

\newcommand{\figref}[1]{\colref{Figure}{#1}}
\newcommand{\tabref}[1]{\colref{Table}{#1}}

\newcommand{\propref}[1]{\colref{Proposition}{#1}}

\newcommand{\logLogSlopeTriangle}[5]
{

    \pgfplotsextra
    {
        \pgfkeysgetvalue{/pgfplots/xmin}{\xmin}
        \pgfkeysgetvalue{/pgfplots/xmax}{\xmax}
        \pgfkeysgetvalue{/pgfplots/ymin}{\ymin}
        \pgfkeysgetvalue{/pgfplots/ymax}{\ymax}

        \pgfmathsetmacro{\xArel}{#1}
        \pgfmathsetmacro{\yArel}{#3}
        \pgfmathsetmacro{\xBrel}{#1-#2}
        \pgfmathsetmacro{\yBrel}{\yArel}
        \pgfmathsetmacro{\xCrel}{\xArel}

        \pgfmathsetmacro{\lnxB}{\xmin*(1-(#1-#2))+\xmax*(#1-#2)} 
        \pgfmathsetmacro{\lnxA}{\xmin*(1-#1)+\xmax*#1} 
        \pgfmathsetmacro{\lnyA}{\ymin*(1-#3)+\ymax*#3} 
        \pgfmathsetmacro{\lnyC}{\lnyA+#4*(\lnxA-\lnxB)}
        \pgfmathsetmacro{\yCrel}{\lnyC-\ymin)/(\ymax-\ymin)} 

        \coordinate (A) at (rel axis cs:\xArel,\yArel);
        \coordinate (B) at (rel axis cs:\xBrel,\yBrel);
        \coordinate (C) at (rel axis cs:\xCrel,\yCrel);

        \draw[#5]   (A)-- node[pos=0.5,anchor=north] {1}
                    (B)-- 
                    (C)-- node[pos=0.5,anchor=west] {#4}
                    cycle;
    }
}

\date{}
\title{\textbf{A Semi-Implicit Variational Multiscale Formulation for the Incompressible Navier–Stokes Equations via Exact Adjoint Linearization}} 
\author[1$\dag$*]{Biswajit Khara}
\author[1$\dag$]{Suresh Murugaiyan}
\author[2]{Suriya Dhakshinamoorthy}
\author[3]{Makrand Khanwale}
\author[2]{Ming-Chen Hsu}
\author[1,2*]{Baskar Ganapathysubramanian}
\affil[1]{\tracisu}
\affil[2]{\meisu}
\affil[3]{\TUM}
\affil[*]{Corresponding author(s). E-mail(s): \href{mailto:bkhara@iastate.edu}{bkhara@iastate.edu}, \href{mailto:baskarg@iastate.edu}{baskarg@iastate.edu}}
\affil[$\dag$]{These authors contributed equally to this work.}

\usetikzlibrary{calc}
\begin{document}
	
\maketitle

\begin{abstract}
A semi-implicit, residual-based variational multiscale (VMS) formulation is developed for the incompressible Navier--Stokes equations. The convection term is linearized using an extrapolated (Oseen-type) convecting velocity, producing a linear advection operator whose adjoint can be written exactly. Because of this exact adjoint, unresolved-scale contributions enter the weak form without spatial derivatives of the fine-scale velocity, thereby eliminating the case-by-case adjustments that often accompany nonlinear residual-based VMS implementations. The formulation is presented for a generalized linear convection operator encompassing the convective, skew-symmetric, and divergence forms. Since the discrete method is linear by construction and monolithic for velocity and pressure, each time step requires only one linear solve, reducing wall-clock time by a factor of $2$ to $5$ relative to fully implicit nonlinear formulations while maintaining comparable accuracy. Temporal convergence is verified, and validation is performed on the lid-driven cavity, flow past a cylinder, turbulent channel flow, and flow over a NACA0012 airfoil at a high Reynolds number, demonstrating the efficiency of the proposed approach on problems of practical scale.
\end{abstract}

\section{Introduction}
\label{sec:intro}
In many important applications, the finite element method (FEM) is a method of choice for solving the incompressible Navier--Stokes equations (NSE), especially for multiphysics couplings (e.g., fluid--structure interaction~\cite{BazilevsBook1}) and flows on complex geometries. Yet, a direct application of continuous Galerkin FEM with equal-order interpolation to the incompressible NSE is well known to be challenging~\cite{donea2003finite,john2016finite}. Two distinct instabilities arise. First, because pressure does not appear in the continuity equation, the discrete problem is a saddle-point system; unless the velocity--pressure spaces satisfy an inf-sup condition, spurious pressure (and velocity) oscillations result \cite{babuvska1971error,brezzi1974existence}. Second, in convection-dominated regimes, the standard Galerkin formulation lacks the necessary numerical dissipation and exhibits nonphysical oscillations \cite{franca2004stabilized}. Additionally, standard formulations also suffer from loss of mass conservation (satisfaction of the continuity equation). These challenges have motivated a broad class of remedies, including nonconforming variational formulations that satisfy continuity by design~\cite{crouzeix1973conforming,reed1973triangular} (these are not inf-sup stable by default), inf-sup stable elements \cite{taylor1973numerical,arnold1984stable} (these still need stabilization terms to satisfy mass conservation), and residual-based stabilizations \cite{brooks1982streamline,hughes1986new,douglas1989absolutely,tezduyar1991stabilized} (these are not inf-sup stable but stabilization terms circumvent the problem).

In this work, we focus on the variational multiscale (VMS) framework \cite{hughes1995multiscale, guermond1999stabilization, hughes2000large, Hughes01b, HugScoFr04, Gravemeier06, bazilevs2007variational, Bazilevs12kn, Takizawa12gp, Takizawa13ik, Bazilevs14dt, Takizawa14dw, Takizawa14hg, Bazilevs15hh, Colomes15, Ahmed17, RasthoferGravemeier18, CodinaEnc18}, which provides a unified stabilization mechanism for both the saddle-point structure and convection-driven instabilities when used with a continuous Galerkin finite element formulation. VMS decomposes the solution into \textit{coarse scales} (resolved by the finite element space) and \textit{fine scales} (unresolved by the mesh). The fine scales are not explicitly represented, but their effect on the resolved scales is incorporated through a closure model that expresses fine-scale dynamics in terms of coarse-scale residuals and stabilization parameters. In classical residual-based VMS, this yields pressure-, upwinding-, and grad/div- stabilizations naturally within a consistent variational setting.

However, when VMS is applied to nonlinear operators (such as the Navier--Stokes equations), it results in nonlinear coupling between the resolved (coarse) and unresolved (fine) scales, and the resulting formulation yields terms involving spatial derivatives of the fine-scale velocity $u^\prime$. These derivatives depend on derivatives of the coarse-scale residual and on spatial variations of the stabilization parameters. 
The closure model for the fine-scale terms in the discrete sense is generally a piecewise (element-wise) approximation, which lacks the regularity for derivatives, especially for continuous Galerkin basis functions. 
The standard remedy is to ``transfer'' derivatives off $u^\prime$ via integration by parts and related identities. However, for nonlinear convection terms, this transfer is typically performed on a case-by-case basis (and often boundary-condition-dependent), leading to formulations that are less general (see \secref{sec:prelim}).
This paper addresses this difficulty by developing a \textit{semi-implicit} VMS formulation in which the advection operator is linearized (Oseen-type) using known information from previous time levels. Linearization has two coupled benefits. First, it naturally yields a \textit{derivative-free unresolved-scale (fine-scale) representation}: by this, we mean a weak formulation in which unresolved-scale contributions appear \emph{without spatial derivatives of $u^\prime$} (and hence without derivatives of residuals or stabilization parameters), so that all fine-scale terms can be expressed using coarse-scale residuals only. This removes the need for additional weakening steps and the associated special treatment of boundary terms. Second, since the resulting scheme is linear at each time step, it requires the solution of \emph{one linear system per time step}, offering substantial computational savings relative to fully implicit nonlinear VMS formulations.

The key technical observation underpinning the derivative-free representation is that when the convection operator is linear, its adjoint can be computed \emph{exactly}. Beyond implementation simplicity, avoiding derivatives of $u'$ is also important for extensibility. Formulations that minimize boundary-condition-sensitive integration-by-parts arguments are easier to integrate with discretizations and geometric technologies that complicate boundary term bookkeeping, such as shifted boundary methods (SBM)~\cite{yang2024optimal,yang2025simulating,yang2025shifted,karki2025direc}.
In such settings, reducing special boundary treatments is particularly valuable.

Building on these ideas, our contributions are:
\begin{enumerate}
\item \textit{Exact-adjoint observation and derivative-free unresolved (fine)-scale representation:} We show that once the advection operator is linearized, its adjoint can be computed \emph{exactly}. Within a residual-based VMS closure, this enables a systematic transfer of derivatives from unresolved-scale terms onto the test functions, yielding a formulation in which the unresolved-scale contributions appear \emph{without spatial derivatives of $u'$} (and hence without derivatives of residuals or stabilization parameters). This eliminates the need for ad hoc weakening steps and reduces sensitivity to boundary-condition-specific integration-by-parts manipulations.

\item \textit{General semi-implicit VMS formulation and convection-form comparison:} Using the above mechanism, we derive a semi-implicit residual-based VMS formulation for the incompressible NSE based on a generalized linear advection operator, and we present and compare three common linear advection forms.

\item \textit{Computational efficiency:} The proposed method is linear by construction and requires only a single linear solve per time step, providing a significant speedup compared to fully implicit nonlinear VMS formulations.

\end{enumerate}
These contributions are supported by numerical experiments: we verify temporal convergence and validate the method on a suite of standard benchmark problems, comparing its accuracy and computational cost with those of fully implicit counterparts.

The remainder of this paper is organized as follows. In \secref{sec:prelim}, we review preliminaries and motivate the current work. In \secref{sec:formulation}, we present the semi-implicit VMS formulation for a general linear convection operator, a formal analysis of energy stability, and some implementation details. We report our numerical results in \secref{sec:results}, and conclude our discussions in \secref{sec:conclusions}.

\section{Preliminaries and motivation}
\label{sec:prelim}
In this section, we give a brief overview of the fully implicit VMS formulation, highlighting the important steps and illustrating the issues encountered. The presentation in this section also lays down the basic terminology and sets the stage for the next section.

\subsection{Notation}
\label{sec:notation}
Throughout this paper, we will denote vector-valued quantities by boldface letters. For example, $\xvec$ given by
\begin{align*}
\xvec = \left[ \xcomp{1}, \xcomp{2}, \ldots, \xcomp{d} \right]^T =
\begin{bmatrix}
\xcomp{1} \\
\xcomp{2} \\
\vdots \\
\xcomp{d}
\end{bmatrix}
\end{align*}
denotes the spatial coordinates where each $\xcomp{i}$, $i = 1, 2,\ldots, d$ are scalars. Similarly,
$\uvec(t,\xvec) = \left[ u_1, u_2, \ldots, u_n \right]^T $ denotes a vector-valued \emph{function} where each $\{\ucomp{i}\}_{i=1}^{n}$ is a function of both time and space, i.e., $\ucomp{i} = \ucomp{i}(t,\xvec)$. When $\uvec$ is the velocity vector, the number of components equal the spatial dimension, i.e., $n=d$.

Given two scalar-valued functions $a(\xvec)$ and $b(\xvec)$ on a domain $\spDom \subset \mathbb{R}^d$, we denote the inner-product between them as
\begin{align*}
\inner{a}{b} := \intSpace a(\xvec)\ b(\xvec) \mbox{d}\xvec.
\end{align*}
For two \emph{vector-valued} functions $\avec(\xvec)$ and $\bvec(\xvec)$, we denote the inner-product in a similar manner as
\begin{align*}
\inner{\avec}{\bvec} := \intSpace \avec(\xvec) \cdot \bvec(\xvec) \mbox{d}\xvec =\intSpace \avec(\xvec)^T\bvec(\xvec) \mbox{d}\xvec = \intSpace \left[ \sum_{i = 1}^{n} a_i(\xvec)\ b_i(\xvec) \right]\mbox{d}\xvec = \sum_{i = 1}^{n} \inner{a_i}{b_i}.
\end{align*}

\subsection{Implicit Galerkin formulation of the Navier--Stokes equations}
\begin{figure}[!htb]
\centering
\begin{tikzpicture}[line cap=round, line join=round]
	
	\coordinate (P1) at (-2.2,  0.0);   
	\coordinate (P2) at ( 1.8, -0.3);   
	
	\fill[gray!12]
	(P1)
	.. controls (-2.0, -0.8) and (-0.8, -2.0) .. ( 0.2, -1.9)
	.. controls ( 1.2, -1.8) and ( 1.0, -0.7) .. (P2)
	.. controls ( 2.6,  0.1) and ( 1.5,  1.8) .. ( 0.0,  2.2)
	.. controls (-1.5,  2.6) and (-2.4,  0.8) .. (P1);
	
	\draw[very thick]
	(P1)
	.. controls (-2.0, -0.8) and (-0.8, -2.0) .. ( 0.2, -1.9)
	.. controls ( 1.2, -1.8) and ( 1.0, -0.7) .. (P2);
	
	\draw[very thick, dash pattern=on 7pt off 3.5pt]
	(P2)
	.. controls ( 2.6,  0.1) and ( 1.5,  1.8) .. ( 0.0,  2.2)
	.. controls (-1.5,  2.6) and (-2.4,  0.8) .. (P1);
	
	\fill (P1) circle (2.5pt);
	\fill (P2) circle (2.5pt);
	
	\node[font=\large] at ( 0.0,  0.3) {$\Omega$};
	\node[font=\large] at ( 0.0, -2.25) {$\Gamma_D$};
	\node[font=\large] at (-1.5,  2.8) {$\Gamma_N$};
	
	\coordinate (NP) at (0.0, 2.2);
	\draw[-{Stealth[length=9pt, width=5.5pt]}, thick]
	(NP) -- ++(0.20, 0.75);
	\node[font=\normalsize] at ($(NP)+(0.46, 0.82)$) {$\hat{n}$};
	
\end{tikzpicture}
\caption{Sketch of a general domain with different conditions applied at different boundaries. Neumann conditions are prescribed on $\Gamma_N$, and Dirichlet conditions on the velocities are prescribed on $\Gamma_{D}$.}
\label{fig:schematic}
\end{figure}

Suppose $\spDom$ is a spatial domain (see \figref{fig:schematic}). The incompressible Navier--Stokes equations in non-dimensional form on $\spDom$ are
\begin{subequations}
\label{eq:nse}
\begin{align}
\label{eq:nse-momentum}
\timederShort \uvec + (\ddiv{\uvec})\uvec + \dvc(\divergence\uvec)\uvec - \visco\laplacian\uvec + \grad p - \fvec &= \mathbf{0}, \\
 \label{eq:nse-continuity}
\divergence\uvec &= 0,
\end{align}
\end{subequations}
along with the boundary conditions
\begin{subequations}
\label{eq:nse-bc}
\begin{align}
\label{eq:nse-bc-d}
\uvec &= \uvec_{D}\ \text{on}\ \Gamma_{D},\\
\label{eq:nse-bc-n}
\normalvec\cdot(-p\mvec{I}+\visco\grad\uvec) &= \mvec{h} \ \text{on}\ \Gamma_{N},
\end{align}
\end{subequations}
and initial conditions
\begin{align}
\label{eq:nse-ic}
\uvec(t = 0, \cdot) = \uvec_0(\cdot).
\end{align}
Here, $\uvec$, $p$ and $\fvec$ are the velocity, pressure, and forcing, respectively. The parameter $\visco$ is the dimensionless viscosity ($\visco=Re^{-1}$). The scalar parameter $s\in\{0,\halfnice,1\}$ is introduced to toggle between three equivalent nonlinear convection representations,
\begin{align}
(\ddiv{\uvec})\uvec + \dvc(\divergence\uvec)\uvec =
\begin{cases}
(\ddiv{\uvec})\uvec, &\text{when } s=0 \quad \text{(convective form)},\\[2pt]
(\ddiv{\uvec})\uvec + \halfnice (\divergence\uvec)\uvec, &\text{when } s=\halfnice \quad \text{(skew-symmetric form)},\\[2pt]
\divergence(\uvec\uvec^T), &\text{when } s=1 \quad \text{(divergence form)},
\end{cases}
\end{align}
using the identity $\divergence(\uvec\uvec^T) = (\ddiv{\uvec})\uvec + (\divergence\uvec)\uvec$.
Although these forms are equivalent at the continuous level under $\divergence\uvec=0$, they lead to different weak forms once modeling and stabilization are introduced, and they lead to different discrete formulations. We retain this parameterization because the semi-implicit development in \secref{sec:formulation} will treat a generalized \emph{linear} convection operator in an analogous way. Equations \eqref{eq:nse}--\eqref{eq:nse-ic} form a system that can be solved numerically. For unique solvability we require $\Gamma_{D}\neq\emptyset$; otherwise the velocity is determined only up to the constants in the kernel of the viscous operator. When $\Gamma_{N}=\emptyset$, the pressure is in addition determined only up to an additive constant and must be fixed separately, for instance by prescribing its value at a point (see \secref{sec:ldc2d}).

We now derive a Galerkin formulation of \eqref{eq:nse}. Define the trial and test spaces
\begin{subequations}\label{eq:bochner-subspaces-modified}
	\begin{align}
		\spaceTcVd
		&:= \Big\{
		\mvec{v} \in L^2(0,T;\mvec{H}^1(\spDom)) :
		\mvec{v}(t) = \uvec_{D}(t) \ \text{on } \Gamma_{D}, 
		\ \text{for a.e. } t\in(0,T)
		\Big\}, \\
		\spaceTcV
		&:= \left\{
		\mvec{v} \in L^2(0,T;\mvec{H}^1(\spDom)) :
		\mvec{v}(t) = \zerovec
		\ \text{on } \Gamma_{D}
		\ \text{for a.e. } t\in(0,T)
		\right\}, \\
		\spaceTcQ
		&:= L^2(0,T;H^1(\spDom)).
	\end{align}
\end{subequations}
We seek $(\uvec,p)\in \spaceTcVd\times\spaceTcQ$ such that, for all $(\vvec,q)\in \spaceTcV\times\spaceTcQ$,
\begin{align}
\intSpace \vvec\cdot \left[ \timederShort \uvec + (\ddiv{\uvec})\uvec + \dvc(\divergence\uvec)\uvec - \visco\laplacian\uvec + \grad p - \fvec \right] \dxv
+ \intSpace q \ \divergence\uvec \ \dxv = 0.
\end{align}

We integrate the viscous and the pressure terms by parts as
\begin{align}
	\intSpace \vvec \cdot \left[- \visco\laplacian\uvec + \grad p \right] \dxv&= \intSpace \visco \grad \vvec : \grad \uvec\ \dxv - \intSpace (\divergence\vvec)\ p \ \dxv - \int_{\Gamma_{N}} \vvec \cdot \left(\normalvec \cdot \left[-p\mvec{I} + \visco\grad\uvec\right]\right)\ds \\
	&= \intSpace \visco \grad \vvec : \grad \uvec\ \dxv - \intSpace (\divergence\vvec)\ p \ \dxv - \int_{\Gamma_{N}} \vvec \cdot \hvec \ \ds,
\end{align}
where the boundary integral survives only on $\Gamma_{N}$ since $\vvec=0$ on $\Gamma_{D}$.
Collating all terms, and using the inner product notation described in \secref{sec:notation}, the Galerkin formulation of \eqref{eq:nse} is given by
\begin{multline} \label{eq:nse-bdf-galerkin}
\inner{\vvec}{\timederShort \uvec} + \inner{\vvec}{\uvec\cdot\grad\uvec} + \dvc \inner{\vvec}{(\divergence\uvec)\uvec} + \visco\inner{\grad \vvec}{\grad\uvec} \\ - \inner{\divergence\vvec}{p} + \inner{q}{\divergence\uvec} - \inner{\vvec}{\fvec} -\inner{\vvec}{\hvec}_{\Gamma_{N}} = 0
\end{multline}
for any $\vvec\in \spaceTcV $ and $q\in \spaceTcQ$.

\paragraph{Lack of stability of the continuous Galerkin method.}
The continuous Galerkin formulation \eqref{eq:nse-bdf-galerkin} of the NSE suffers from two major issues. The first issue relates to the saddle-point nature of the NSE, i.e., the absence of the pressure variable in the continuity equation. Because of this, only some special pairs of velocity and pressure functions are allowed. These pairs are called \textit{inf-sup stable} pairs. However, a standard implementation of the Galerkin method using equal-order polynomials for both velocities and the pressure clearly violates this, and therefore, a stable numerical solution cannot be guaranteed \cite{babuvska1973finite, brezzi1974existence, john2016finite}.

The second issue relates to the presence of the advective term $(\ddiv{\uvec})\uvec$. The advection term has directional properties, and a continuous Galerkin formulation of this term is known to be unstable when advection dominates (compared to diffusion) \cite{brooks1982streamline, hughes1989new, codina1998comparison}.

There are multiple ways to resolve these two issues. The issue of instabilities due to non-inf-sup conformity can be resolved by using methods such as PSPG \cite{hughes1986new, douglas1989absolutely, tezduyar1992incompressible} that remove the saddle point nature of the resulting FEM problem. The second issue can be resolved by using an ``upwinding''-type modification to the test function $ \vvec $ (e.g., SUPG \cite{brooks1982streamline}) or by using a least squares modification of the FEM problem (e.g., GLS \cite{hughes1989new}). In this work, we apply the variational multiscale method (VMS \cite{hughes1995multiscale}) which provides both kinds of stability.

\subsection{A review of the variational multiscale stabilization}
\label{sec:stabilized-formulation-nse}

The variational multiscale (VMS) method naturally addresses both instabilities discussed above in a consistent manner \cite{bazilevs2007variational}. Assume $\spaceTcVhd \subset \spaceTcVd$, $\spaceTcVh \subset \spaceTcV$ and $\spaceTcQh \subset \spaceTcQ$ are suitable discrete function spaces based on the given mesh. Then, following \cite{hughes1995multiscale, bazilevs2007variational}, we decompose the velocity and pressure as
\begin{subequations}\label{eq:vms-coarse-fine-decomposition}
\begin{align}
\uvec(\xvec) &= \uvech(\xvec) + {\uf}(\xvec), \\
p(\xvec) &= \ph(\xvec) + {\pf}(\xvec).
\end{align}
\end{subequations}
Here, $(\uvech,\ph) \in \spaceTcVhd \times \spaceQh$ denote the \emph{coarse scales} represented by the finite element spaces, while $(\uf,\pf)$ denote the \emph{fine scales} unresolved by the mesh. In residual-based VMS, the fine scales are modeled \emph{element-wise} in terms of the coarse-scale residuals \cite{hughes1995multiscale, bazilevs2007variational} as
\begin{subequations}
\label{eq:nse-fine-scale-approximations}
\begin{align}
\uf &\approx \uprime := -\taum \resMom, \\
\pf &\approx \pprime := -\tauc \resCon,
\end{align}
\end{subequations}
where the element-wise strong-form residuals are
\begin{subequations}
\label{eq:coarse-scale-residuals}
\begin{align}
\resMom &:= \timederShort\uvech + \uvech\cdot\grad\uvech + \dvc(\divergence\uvech)\uvech -\visco\laplacian\uvech + \grad\ph - \fvec, \\
\resCon &:= \divergence\uvech.
\end{align}
\end{subequations}
The stabilization parameters $\taum>0$ and $\tauc>0$ are defined element-wise following \cite{shakib1991new, bazilevs2007variational, Hsu09a, Takizawa17e} (see also Remark~\ref{rem:tau-def} and Remark~\ref{rem:tau-uniform}).

Substituting \eqref{eq:nse-fine-scale-approximations} into \eqref{eq:nse-bdf-galerkin}, and restricting the test functions to the coarse-scale test space yields
\begin{multline}
\label{eq:nse-vms}
    \inner{\vvech}{\timederShort(\uvech+\uprime)} + \inner{\vvech}{(\uvech+\uprime)\cdot\grad(\uvech+\uprime)} + s \inner{\vvech}{(\divergence(\uvech+\uprime))(\uvech+\uprime)} \\
    + \visco\inner{\grad \vvech}{\grad(\uvech+\uprime)} - \inner{\divergence\vvech}{\ph+\pprime}
    + \inner{\qh}{\divergence(\uvech+\uprime)} - \inner{\vvech}{\fvec} -\inner{\vvech}{\hvec}_{\Gamma_{N}} = 0
\end{multline}
for all $(\vh,\qh) \in \spaceTcVh\times\spaceQh$. This is the full nonlinear VMS formulation of \eqref{eq:nse}. Since $(\uprime,\pprime)$ are defined element-wise, all terms involving $\uprime$ and $\pprime$ in \eqref{eq:nse-vms} are to be interpreted as element-wise integrals over the mesh. Since $(\uprime,\pprime)$ are expressed in terms of $(\uvech,\ph)$, the only global unknowns are the coarse scales.

\begin{remark}[Stabilization parameters]
	\label{rem:tau-def}
	Once a full discretization in space and time is chosen, the stabilization parameters can be defined as below. Let $\mvec{\xi}\mapsto\xvec$ denote the map from the reference element onto $K\in\mesh$, and let $\mmat{G}$ and $\mvec{g}$ denote the element metric tensor and metric vector,
	\begin{align}
		\label{eq:metric-tensor}
		G_{ij} := \sum_{k}\partialder{\xi_k}{x_i}\,\partialder{\xi_k}{x_j},
		\qquad
		g_{i} := \sum_{j}\partialder{\xi_j}{x_i} .
	\end{align}
	Then following \cite{shakib1991new, bazilevs2007variational, Hsu09a, Takizawa17e}, and assuming a BDF discretization in time, the stabilization parameters are defined as
	\begin{subequations} \label{eq:taum-tauc-shakib}
		\begin{align}
			\taum &= \left[\, \left(\frac{2 \beta_0}{\dt}\right)^2
			+ \avech\cdot\mmat{G}\,\avech
			+ \cinv\,\visco^2\,\mmat{G}\!:\!\mmat{G} \,\right]^{-\halfnice},
			\label{eq:taum-metric}\\
			\tauc &= \frac{1}{\taum\,\left(\mvec{g}\cdot\mvec{g}\right)} .
			\label{eq:tauc-metric}
		\end{align}
	\end{subequations}
	Here, $\dt$ is the time-step, and $\beta_0$ is the leading BDF coefficient (see \secref{sec:time-disc}). The second term is related to the advection, and $\avech$ is an estimate of the local advective field. We set $\avech=\uvech$ in the nonlinear formulation \eqref{eq:nse-vms}, and the extrapolated field of \secref{sec:adv-field-spec} in the semi-implicit formulation \eqref{eq:oseen-vms-bdf}. The constant $\cinv>0$ arises from an element-wise inverse estimate; we use the standard values $\cinv=36$ for linear and $\cinv=60$ for quadratic elements \cite{bazilevs2007variational}.
\end{remark}

\begin{remark}[Uniform-mesh form]
	\label{rem:tau-uniform}
	For a uniform, axis-aligned element of size $h$ in $d$ space dimensions, with reference element $[-1,1]^d$, one has $\mmat{G}=(4/h^2)\,\mmat{I}$ and $\mvec{g}\cdot\mvec{g}=4d/h^2$, so \eqref{eq:taum-tauc-shakib} reduces to the scalar form
	\begin{align}
		\label{eq:tau-uniform}
		\taum = \left[\,\frac{c_1}{\dt^2}
		+ \frac{c_2\,|\avech|^2}{h^2}
		+ \frac{c_3\,\cinv\,\visco^2}{h^4}\,\right]^{-\halfnice},
		\qquad
		\tauc = \frac{h^2}{c_4\,\taum} ,
	\end{align}
	where $c_1 = 4\beta_0^2$, $c_2 = 4$, $c_3 = 16d$ and $c_4 = 4d$.
\end{remark}

\subsubsection{Derivatives of the fine scale velocities}
\label{sec:vms-derivatives-of-fine-scale}

Equation \eqref{eq:nse-vms} involves spatial derivatives of the modeled fine-scale velocity $\uprime$. This is problematic because differentiating $\uprime=-\taum\resMom$ introduces derivatives of both the residual and the stabilization parameter. For example,
\begin{align}
\partialder{\uprime}{x_i}
 = \partialder{}{x_i}(-\taum \resMom)
 = -\left[ \resMom \partialder{\taum}{x_i} + \taum \partialder{\resMom}{x_i}\right].
\end{align}
Assuming a uniform mesh, the derivative of $\taum$ can be expressed as
\begin{align}
	\label{eq:dtaum-dx}
	\partialder{\taum}{x_i}
	= -\half \left[ \frac{c_1}{\dt^2} + \frac{c_2 |\uvech|^2}{h^2} + \frac{c_3\cinv\visco^2}{h^4} \right]^{-\nicefrac{3}{2}} \times
	\left[\frac{2c_2}{h^2}\,\uvech\cdot\partialder{\uvech}{x_i}\right]
	= -\frac{c_2}{h^2}\,\taum^{3}\,\uvech\cdot\partialder{\uvech}{x_i},
\end{align}
which is already nonlinear in $\uvech$. The second term contains
\begin{align}
\partialder{\resMom}{x_i}
= \partialder{}{x_i}\left[\timederShort\uvech + \uvech\cdot\grad\uvech + \dvc(\divergence\uvech)\uvech -\visco\laplacian\uvech + \grad\ph - \fvec\right],
\end{align}
which involves third derivatives of $\uvech$, second derivatives of $\ph$, and derivatives of $\fvec$. These differentiability requirements exceed those of the original weak form and are generally incompatible with low-order continuous finite element spaces. For this reason, practical residual-based VMS implementations typically avoid explicit differentiation of $\resMom$ and $\taum$. The remaining difficulty is then how to rewrite the weak form so that $\uprime$ appears without derivatives.

In \eqref{eq:nse-vms}, the time derivative of $\uprime$ appears in the first term, and spatial derivatives appear in the continuity term, the diffusion term, and the convection term. The standard treatments for these terms are as below:
\begin{itemize}
	\item \textbf{The acceleration term.} We assume that the projection of $\timederShort\uprime$ on to $\spaceVh$ is negligible (see, e.g., \cite{bazilevs2007variational}), therefore we write
	\begin{align*}
		\inner{\vvech}{\timederShort(\uvech + \uvec')} &= \inner{\vvech}{\timederShort\uvech} + \inner{\vvech}{\timederShort\uvec'} \\
		&= \inner{\vvech}{\timederShort \uvech}.
	\end{align*}
	We note that there exist VMS formulations that do not neglect the fine-scale time derivatives, see for instance \cite{codina2002stabilized, CoPrGuBa07, GaGrWa10}.
	
	\item \textbf{Diffusion term:} The diffusion contribution produces $\inner{\grad \vvech}{\grad \uprime}$, for which transferring derivatives off $\uprime$ is not straightforward in low-order settings. Following \cite{bazilevs2007variational}, we neglect this term, i.e.,
	\begin{align*}
		\inner{\grad \vvech}{\grad(\uvech+\uprime)} \approx \inner{\grad \vvech}{\grad\uvech}.
	\end{align*}
	Since the Laplacian with Dirichlet conditions is coercive, this modification is commonly justified by viewing $\uvech$ as a Riesz projection of $\uvec$, so that omitting $\inner{\grad \vvech}{\grad \uprime}$ does not compromise diffusion stability \cite{john2016finite}.

\item \textbf{Continuity term:} The term $\inner{\qh}{\divergence(\uvech+\uprime)}$ can be treated by integration by parts, which transfers derivatives onto the test function:
\begin{align*}
\inner{\qh}{\divergence(\uvech+\uprime)}
&= \inner{\qh}{\divergence\uvech} + \inner{\qh}{\divergence\uprime}
= \inner{\qh}{\divergence\uvech} - \inner{\grad \qh}{\uprime} + \innerB{\qh}{\normalvec\cdot\uprime}.
\end{align*}

\item \textbf{Advection term:} The convection contribution is more delicate: it cannot be discarded without affecting stability, and must be handled on a case-by-case basis. Here, we discuss this for $\dvc=0$ and $\dvc=1$.
\begin{enumerate}
\item \textit{Case $s=1$ (divergence form):}
Using integration by parts,
\begin{multline*}
\inner{\vvech}{\divergence\!\left((\uvech+\uprime)(\uvech+\uprime)^T\right)}
= - \inner{\grad \vvech}{(\uvech+\uprime)(\uvech+\uprime)^T}
 + \\ \innerB{\vvech}{\normalvec\cdot \left[ (\uvech+\uprime) (\uvech+\uprime)^T \right]}.
\end{multline*}

\item \textit{Case $s=0$ (convective form):} A common manipulation is to add and subtract a term based on the modeling interpretation that $(\uvech+\uprime)$ represents the full velocity and should therefore be (approximately) solenoidal. Specifically, one introduces $\divergence(\uvech+\uprime)\,\uprime$ under the integral to rewrite
\begin{align}
\inner{\vvech}{(\uvech+\uprime)\cdot\grad(\uvech+\uprime)}
&= \inner{\vvech}{(\uvech+\uprime)\cdot\grad(\uvech+\uprime)} + \inner{\vvech}{\divergence(\uvech+\uprime)\ \uprime} \nonumber\\
 &= \inner{\vvech}{\ddiv{(\uvech+\uprime)}\uvech} + \inner{\vvech}{\divergence \left[\uprime(\uvech+\uprime)^T\right]} \nonumber\\
 &= \inner{\vvech}{\ddiv{(\uvech+\uprime)}\uvech} - \inner{\grad \vvech}{\uprime(\uvech+\uprime)^T} \nonumber \\ &\hspace{1.6in}
 + \innerB{\vvech}{\normalvec\cdot \left[ \uprime(\uvech+\uprime)^T \right]} \nonumber.
 \end{align}
\end{enumerate}
\end{itemize}

\begin{remark}
The two convection treatments differ by the volume term $\inner{\vvech}{(\divergence(\uvech+\uprime))\,\uvech}$ and by the form of the resulting boundary flux contributions. While these boundary terms vanish under full Dirichlet conditions on $\partial\spDom$, they do not generally vanish when mixed boundary conditions are present. Consequently, the weakening steps used to avoid derivatives of $\uprime$ can be sensitive to the boundary partition and to how boundary integrals are treated numerically.
\end{remark}

\section{Linear semi-implicit variational formulation}
\label{sec:formulation}

In this section, we present a linear semi-implicit stabilized method based on the variational multiscale framework. This has two main advantages: (i) it leads to a simpler and unified formulation of the three convection forms, and (ii) it is faster than a full nonlinear method. The key device is to replace the nonlinear convection velocity by a known convecting field, leading to an Oseen-type problem. For this linear convection operator, an exact adjoint can be written, enabling a systematic transfer of derivatives onto test functions.

\subsection{Oseen equations with generalized advection}

The Oseen equations are a linear approximation of the Navier--Stokes equations in which the convecting velocity is prescribed. In the Navier--Stokes equations \eqref{eq:nse}, the unknown velocity $\uvec$ is transported by itself. The Oseen approximation replaces this convecting velocity by a known field $\avec$, yielding a linear problem for $(\uvec,p)$:
\begin{subequations}
\label{eq:oseen}
\begin{align}
\label{eq:oseen-momentum}
\timederShort \uvec + \opAdvGen[\avec]{s}\uvec - \visco\laplacian\uvec + \grad p - \fvec &= \zerovec, \\
\label{eq:oseen-continuity}
\divergence\uvec &= 0,
\end{align}
\end{subequations}
with the same boundary conditions as in \eqref{eq:nse-bc}. Here $s\in\{0,\tfrac{1}{2},1\}$ parameterizes three common linear advection forms. We define the generalized linear advection operator
\begin{align}
\label{eq:advection-op-general}
\opAdvGen[\avec]{s}\uvec := (\avec\cdot\nabla)\uvec + \dvc(\divergence\avec)\uvec.
\end{align}
For $s=0,\sfrac12,1$, this recovers the convective, skew-symmetric, and divergence forms, summarized in \tabref{tab:convection-operators}. Unlike the nonlinear operator $\opAdvGen[\uvec]{s}\uvec$, the operator \eqref{eq:advection-op-general} is linear in $\uvec$ and admits an exact adjoint in the weak form, which is central to what follows. The concrete specification of $\avec$ is provided in \secref{sec:adv-field-spec}.
\begin{table}[!htb]
	\centering
	\caption{$\opAdvGen{\dvc}\uvec = (\adiv)\uvec + \dvc(\divergence\avec)\uvec $, for three chosen values of $\dvc$.}
	\label{tab:convection-operators}
	\begin{tabular}{@{}c|c|c|cc@{}}
		\toprule[2pt]
		\multirow{2}{*}{$\dvc$} &
		\multirow{2}{*}{\begin{tabular}[c]{@{}c@{}}Operator\\ symbol\end{tabular}} &
		\multirow{2}{*}{\begin{tabular}[c]{@{}c@{}}Descriptive\\ name\end{tabular}} &
		\multicolumn{2}{c}{Expression} \\ \cmidrule(l){4-5} 
		& & & \multicolumn{1}{c|}{Vector notation}& {Index notation}\\ \midrule[2pt]
		0 & $\opCAdv := \opAdvGen{0}$ & \begin{tabular}[c]{@{}c@{}}Convective\\ form\end{tabular} & \multicolumn{1}{c|}{$\opConvForm{\uvec}$} & $(a_j\partial_j)u_i$ \\ \midrule
		$\halfnice$ &
		$\opSAdv := \opAdvGen{\halfnice}$ &
		\begin{tabular}[c]{@{}c@{}}Skew-symmetric\\ form\end{tabular} &
		\multicolumn{1}{c|}{$\opSkewForm{\uvec}$} &
		$(a_j\partial_j)\ui + \halfnice (\partial_j a_j)\ui $ \\ \midrule		
1 & $\opDAdv := \opAdvGen{1}$ & \begin{tabular}[c]{@{}c@{}}Divergence\\ form\end{tabular} & \multicolumn{1}{c|}{$\opDivForm{\uvec}$}& $\partial_j(u_i a_j)$ \\ \bottomrule[2pt]
	\end{tabular}
\end{table}

\subsubsection{Exact adjoint of the Oseen advection}
The operator $\opAdvGen[\avec]{s}$ has the property that, when tested against a sufficiently smooth $\vvec$, all derivatives acting on $\uvec$ can be transferred to $\vvec$ exactly.

\begin{proposition}(Exact adjoint relation for $\opAdvGen[\avec]{s}$) \label{thm:adv-adjoint}
Let $\avec$ be sufficiently smooth and let $\uvec,\vvec$ be sufficiently smooth vector fields with well-defined traces on $\partial\Omega$. Then
\begin{align}
\label{eq:adv-weaken-relation}
\inner{\vvec}{\opAdvGen[\avec]{s}\uvec}
= \inner{-\opAdvGen[\avec]{1-s}\vvec}{\uvec} + \innerB{\vvec}{\normalvec\cdot(\uvec\avec^T)}.
\end{align}
\end{proposition}

\begin{proof}
Using $\divergence(\uvec\avec^T)=(\avec\cdot\nabla)\uvec+(\divergence\avec)\uvec$, we write
\begin{align*}
\inner{\vvec}{\opAdvGen[\avec]{s}\uvec}
&= \inner{\vvec}{(\avec\cdot\nabla)\uvec + \dvc(\divergence\avec)\uvec} \\
&= \inner{\vvec}{\divergence(\uvec\avec^T)} + (s-1)\inner{\vvec}{(\divergence\avec)\uvec} \\
&= -\inner{\grad\vvec}{\uvec\avec^T} + \innerB{\vvec}{\normalvec\cdot(\uvec\avec^T)} + (s-1)\inner{(\divergence\avec)\vvec}{\uvec}\\
&= -\inner{(\avec\cdot\nabla)\vvec}{\uvec} - (1-s)\inner{(\divergence\avec)\vvec}{\uvec}
+ \innerB{\vvec}{\normalvec\cdot(\uvec\avec^T)}\\
&= \inner{-\opAdvGen[\avec]{1-s}\vvec}{\uvec} + \innerB{\vvec}{\normalvec\cdot(\uvec\avec^T)}.
\end{align*}
\end{proof}

\begin{remark}
\propref{thm:adv-adjoint} is the key mechanism that removes case-by-case weakening for the \emph{advection} contribution in residual-based VMS: once convection is linearized, the exact adjoint transfers derivatives onto the test functions in a uniform way for all $s\in\{0,\tfrac12,1\}$. In particular, convection-related terms can be written without requiring spatial derivatives of the modeled fine-scale velocity $u'$.
\end{remark}
\begin{remark}
If we take $\vvec$ to be an infinitely smooth function with compact support in $\spDom$ (i.e., $\vvec$ and its derivatives vanish near $\partial\spDom$), then the boundary term in \eqref{eq:adv-weaken-relation} vanishes, and we have
\begin{align}
\inner{\vvec}{\opAdvGen{\dvc} \uvec} = -\inner{\opAdvGen{1-\dvc} \vvec}{\uvec}.
\end{align}
The operator acting on $\vvec$ is the adjoint operator of $\opAdvGen{\dvc}$, and is denoted by $\opAdvGenStar{\dvc}$. So, we have
\begin{align*}
\opAdvGenStar{\dvc} =-\opAdvGen{1-\dvc}.
\end{align*}
\tabref{tab:adjoint-operator-summary} summarizes the adjoint expressions for each value of $\dvc\in \{ 0, \halfnice, 1 \}$.
\end{remark}

\begin{table}[!htb]
	\centering
	\caption{Summary of $\opAdvGen{\dvc}$ and $\opAdvGenStar{\dvc}$}
	\label{tab:adjoint-operator-summary}
	\begin{tabular}{@{}c|c|c|c@{}}
		\toprule [2pt]
		$\dvc$& \begin{tabular}[c]{@{}c@{}}Descriptive\\ name\end{tabular}& \begin{tabular}[c]{@{}c@{}}Operator\\ expression \\ $\opAdvGen{\dvc}\uvec$ \end{tabular} & \begin{tabular}[c]{@{}c@{}}Adjoint\\ expression \\ $\opAdvGenStar{\dvc}\vvec$ \end{tabular} \\ \midrule[2pt]
		0 & \begin{tabular}[c]{@{}c@{}}Convective\\ form\end{tabular} & $\opConvForm{\uvec}$ & $-\opDivForm{\vvec}$ \\ \midrule
		$\halfnice$ & \begin{tabular}[c]{@{}c@{}}Skew-symmetric\\ form\end{tabular} & $\opSkewForm{\uvec}$ & $-\left[ \opSkewForm{\vvec} \right]$ \\ \midrule
		1 & \begin{tabular}[c]{@{}c@{}}Divergence\\ form\end{tabular} & $\opDivForm{\uvec}$ & $-\opConvForm{\vvec}$ \\ \bottomrule[2pt]
	\end{tabular}
\end{table}

\subsection{Stabilized formulation of the Oseen equations}
\label{sec:stabilized-formulation-oseen}

The Oseen system \eqref{eq:oseen} exhibits the same saddle-point and convection-driven instabilities as the nonlinear NSE \cite{BraackBurman06, BraackBurmanJohnLube07, deFrutos16}, so we again apply residual-based VMS. The Galerkin weak form of \eqref{eq:oseen} reads: given $\avec\in \spaceTcVd$, find $(\uvec,p)\in \spaceTcVd\times\spaceTcQ$ such that for all $(\vvec,q)\in \spaceTcV\times\spaceTcQ$,
\begin{align}
\label{eq:oseen-galerkin-1}
\inner{\vvec}{\timederShort \uvec} + \inner{\vvec}{\opAdvGen[\avec]{s}\uvec} + \visco\inner{\grad \vvec}{\grad\uvec}
- \inner{\divergence\vvec}{p} + \inner{q}{\divergence\uvec}
- \inner{\vvec}{\fvec} -\inner{\vvec}{\hvec}_{\Gamma_{N}} = 0.
\end{align}

We now apply the VMS decomposition
\begin{subequations}
\label{eq:oseen-vms-coarse-fine-decomposition}
\begin{align}
\uvec &= \uvech + \uf, \\
p &= \ph + \pf,
\end{align}
\end{subequations}
and model the fine scales element-wise as
\begin{subequations}
\label{eq:oseen-fine-scale-approximations}
\begin{align}
\uf &\approx \uprime := -\taum \resMom (\uvech, \ph), \\
\pf &\approx \pprime := -\tauc \resCon (\uvech),
\end{align}
\end{subequations}
with element-wise residuals of the Oseen strong form
\begin{subequations}
\label{eq:oseen-coarse-scale-residuals}
\begin{align}
\resMom(\uvech, \ph) &:= \timederShort\uvech + \opAdvGen[\avec]{s}\uvech -\visco\laplacian\uvech + \grad\ph - \fvec, \\
\resCon(\uvech) &:= \divergence\uvech.
\end{align}
\end{subequations}
Substituting the decomposition into \eqref{eq:oseen-galerkin-1} gives
\begin{multline}
\label{eq:oseen-vms-modeled}
\inner{\vvec}{\timederShort (\uvech+\uprime)} + \inner{\vvec}{\opAdvGen[\avec]{s}(\uvech+\uprime)}
+ \visco\inner{\grad \vvec}{\grad(\uvech+\uprime)} - \inner{\divergence\vvec}{\ph+\pprime} \\
+ \inner{q}{\divergence(\uvech+\uprime)}
- \inner{\vvec}{\fvec} -\inner{\vvec}{\hvec}_{\Gamma_{N}} = 0.
\end{multline}
Similar to \secref{sec:vms-derivatives-of-fine-scale}, we neglect the following:
\begin{myitemize}
	\item The rate of change of $\uvec'$ in the acceleration term, i.e., we set $\inner{\vvec}{\timederShort \uvec'} = 0$.
	\item The $H^1$-projection of $\uvec'$ onto $\spaceVh$, i.e., $ \inner{\grad \vvec}{\grad \uvec'} = 0$.
	\item The fine scales on all element boundaries, i.e., $\uprime |_{\partial K} = \zerovec$ for every $K\in\mesh$ \cite{hughes1995multiscale, bazilevs2007variational}; in particular $\uprime=\zerovec$ on $\partial\spDom$.
	
\end{myitemize}
And additionally:
\begin{myitemize}
	\item We set the advective field $\avec$ to be a coarse-scale quantity, i.e., we neglect the fine scales of the prescribed advection. So, we have $\avec = \avech$ or $\avec' = 0$.
\end{myitemize}
Then restricting test functions to the coarse-scale test spaces $(\vvech,\qh)$ yields the stabilized semi-implicit Oseen--VMS formulation:
\begin{multline}
\label{eq:oseen-vms}
\inner{\vvech}{\timederShort \uvech} + \inner{\vvech}{\opAdvGen[\avech]{s}\uvech}
+ \visco\inner{\grad \vvech}{\grad\uvech} - \inner{\divergence\vvech}{\ph} + \inner{\qh}{\divergence\uvech} \\
- \inner{\opAdvGen[\avech]{1-s}\vvech}{\uprime} -\inner{\grad\qh}{\uprime}
- \inner{\divergence\vvech}{\pprime}
- \inner{\vvech}{\fvec} -\inner{\vvech}{\hvec}_{\Gamma_{N}} = 0.
\end{multline}
Here, we have used
\begin{align*}
	\inner{\vvech}{\opAdvGen[\avech]{s}\uprime} = - \inner{\opAdvGen[\avech]{1-s}\vvech}{\uprime},
\end{align*}
(see \propref{thm:adv-adjoint}), applied element-wise: the fine-scale terms are element-wise integrals, so the transfer of derivatives produces boundary contributions on each $\partial K$, and these vanish because $\uprime|_{\partial K}=\zerovec$.

Equation \eqref{eq:oseen-vms} is the time-continuous VMS-stabilized formulation of the Oseen approximation \eqref{eq:oseen}. Thus, \eqref{eq:oseen-vms} can be viewed as a linear counterpart of \eqref{eq:nse-vms}. Two features distinguish it from the fully implicit nonlinear VMS form. First, linearization of the convection operator makes its adjoint available in closed form, so the convection contribution can be written in a uniform weak form for all $s\in\{0,\tfrac12,1\}$ without case-by-case weakening of fine-scale convection terms. Second, the resulting time-marching method is linear at each step and therefore requires only one linear solve per time step, which yields substantial savings relative to fully implicit nonlinear VMS formulations.

\subsubsection{Time discretization}
\label{sec:time-disc}

We have kept the time-discretization unspecified so far. Eq. \eqref{eq:oseen-vms} can be solved with various implicit methods, such as Crank--Nicolson, the BDF family and implicit Runge--Kutta methods. In this work, we use the BDF2 scheme for time-discretization. An $r$-th order BDF approximation of $\partial\uvec/\partial t$ at time step $n$ is
\begin{align}
	\label{eq:bdf-disc}
	\partialder{\uvech}{t}\Big|_{n}
	= \oneOver{\dt}\sum_{i = 0}^{r} \beta_{-i}\uvech^{n-i}
	= \sigma \uvech^{n} + \oneOver{\dt}\sum_{i=1}^{r}\beta_{-i}\uvech^{n-i},
\end{align}
where $\sigma=\beta_0/\dt$ and $\{\beta_{-i}\}$ depend on $r$. In a time-marching setting, $\uvech^n$ is unknown, while $\uvech^{n-1},\ldots,\uvech^{n-r}$ are known.

The discrete fine scales $(\uprime,\pprime)$ at the $n$-th time step can now be written as
\begin{subequations}
	\label{eq:oseen-timedisc-scale-residuals}
	\begin{align}
		\uprime &= -\taum \resMom (\uvechN, \phN) = -\taum \left[\sigma\uvechN + \opAdvGen[\avech]{s}\uvechN -\visco\laplacian\uvechN + \grad\phN - \gvecN\right], \\
		\pprime &= -\tauc \resCon (\uvechN) = -\tauc \divergence\uvechN,
	\end{align}
\end{subequations}
where we have defined
\begin{align}
	\gvecN := \fvecN - \oneOver{\dt}\sum_{i=1}^{r}\beta_{-i}\uvech^{n-i}.
\end{align}
Let us define the following function spaces for the \emph{time-discrete} trial and test functions.
\begin{subequations}\label{eq:fem-subspaces-timediscrete}
	\begin{align}
		\spaceVd &:= \left\{ \mvec{v} \in \mvec{H}^1(\spDom): \vvec = \uvec_{D} \ \text{on}\ \Gamma_{D} \right\}, \\
		\spaceV &:= \left\{ \mvec{v} \in \mvec{H}^1(\spDom): \mvec{v} = \zerovec \text{ on } \Gamma_{D}\right\}, \\
		\spaceQ &:= {H}^1(\spDom),
	\end{align}
\end{subequations}
and let us denote the respective finite dimensional spaces by $\spaceVhd$, $\spaceVh$ and $\spaceQh$, based on a finite element mesh given by $\mesh$. Then substituting \eqref{eq:bdf-disc} and \eqref{eq:oseen-timedisc-scale-residuals} in \eqref{eq:oseen-vms}, we have the BDF-$r$ discretization of \eqref{eq:oseen-vms}: given $\avech\in \spaceVhd$, find $(\uvechN, \phN)\in \spaceVhd\times\spaceQh$ such that
\begin{empheq}[box=\fbox]{multline}
	\label{eq:oseen-vms-bdf}
	\inner{\vvech}{\sigma \uvechN} + \inner{\vvech}{\opAdvGen[\avech]{s}\uvechN}
	+ \visco\inner{\grad \vvech}{\grad\uvechN} - \inner{\divergence\vvech}{\phN} + \inner{\qh}{\divergence\uvechN} \\ + \sum_{K \in \mesh} \inner{\divergence\vvech}{\tauc \divergence \uvechN}_K 
	+ \sum_{K \in \mesh} \inner{\opAdvGen[\avech]{1-s}\vvech}{\taum\left[\sigma\uvechN + \opAdvGen[\avech]{s}\uvechN -\visco\laplacian\uvechN + \grad\phN\right]}_K \\
	+ \sum_{K \in \mesh} \inner{\grad\qh}{\taum\left[\sigma\uvechN + \opAdvGen[\avech]{s}\uvechN -\visco\laplacian\uvechN + \grad\phN\right]}_K
	\\
	= \inner{\vvech}{\gvecN} + \sum_{K \in \mesh} \inner{\taum \opAdvGen[\avech]{1-s}\vvech + \taum \grad \qh}{\gvecN}_K +\inner{\vvech}{\hvec}_{\Gamma_{N}}
\end{empheq}
for $(\vvech, \qh) \in \spaceVh\times \spaceQh$.

\begin{remark}[Other Petrov--Galerkin formulations]
	If we replace $\opAdvGen[\avech]{1-s}\vvech$ by $\opAdvGen[\avech]{s}\vvech$ in \eqref{eq:oseen-vms-bdf}, we obtain the standard SUPG/PSPG formulation of \eqref{eq:oseen} with BDF-$r$ time discretization.
\end{remark}
\subsubsection{The advective field}
\label{sec:adv-field-spec}
Once a time-discretization is chosen, we can then specify the known advective field $\avec$ based on the history of the velocity field. In particular, we use an extrapolation of the velocity field from previously computed coarse-scale velocities. Since we neglect $\avec'$ in \eqref{eq:oseen-vms-bdf}, we only need to specify $\avech$.

Suppose the discrete initial condition for the velocity is given by $\uvech^0$. In the BDF2 scheme, the first step is taken using BDF1 (or backward Euler). So, for the first step, $n=1$, the history variable is simply $\uvech^0$. For $n\geq 2$, the (minimum) required velocity history consists of $\uvech^{n-1}$ and $\uvech^{n-2}$. Therefore, we can specify the advective field $\avech$ according to the scheme and the history, i.e.,
\begin{align}
	\label{eq:a-extrap}
	\avech =
	\begin{cases}
		\uvech^{n-1}, &(\text{first-order extrapolation at } n=1),\\
		2\uvech^{n-1}-\uvech^{n-2}, &(\text{second-order extrapolation for } n\geq 2).
	\end{cases}
\end{align}

\begin{remark}[Implications of neglecting $\avec'$]
    Using \eqref{eq:advection-op-general} with the VMS decomposition of $\uvec$, we have
	\begin{align}
		- \inner{\opAdvGen[\avec]{1-\dvc}\vvech}{(\uvechN+\uprime)}
		= \underbrace{- \inner{(\adiv)\vvech}{(\uvechN+\uprime)}}_{\text{(I)}}
		\;\underbrace{-\,(1-\dvc)\inner{(\divergence\avec)\vvech}{(\uvechN+\uprime)}}_{\text{(II)}}.
	\end{align}
	Now, term (I) is independent of $\dvc$. Setting $\avec = \avech + \avec'$, we have:
	\begin{align}
		\text{(I)}
		&= - \inner{\grad \vvech}{(\uvechN+\uprime)\, (\avech + \avec')^T} \nonumber \\
		&= - \inner{\grad \vvech}{\uvechN\avech^T + \uvechN(\avec')^T + \uvec'\avech^T + \uvec'(\avec')^T}.
	\end{align}
	Whereas, term (II) is nonzero only for $\dvc \neq 1$, and expands to
	\begin{align}
		\text{(II)} = -(1-\dvc)\Big[
		&\inner{(\divergence\avech)\vvech}{\uvechN}
		+ \inner{(\divergence\avech)\vvech}{\uprime} \nonumber \\
		&+ \inner{(\divergence\avec')\vvech}{\uvechN}
		+ \inner{(\divergence\avec')\vvech}{\uprime} \Big].
	\end{align}
	When $\avec'$ is neglected, every term containing $\avec'$ drops in both (I) and
	(II), leaving
	\begin{align}
		- \inner{\grad \vvech}{\uvechN\avech^T + \uvec'\avech^T}
		- (1-\dvc)\Big[ \inner{(\divergence\avech)\vvech}{\uvechN}
		+ \inner{(\divergence\avech)\vvech}{\uprime} \Big].
	\end{align}
    In particular, the Reynolds stress-like term $\uvec'(\avec')^T$ vanishes for every $\dvc\in\{0,\halfnice,1\}$.
\end{remark}

\begin{remark}[Retaining $\avec'$]
	The convective term in \eqref{eq:oseen-vms-modeled} is given by
	\begin{align}
		- \inner{\opAdvGen[\avec]{1-\dvc}\vvech}{(\uvechN+\uprime)} &= - \inner{\left[(\avec\cdot\grad)\vvech + (1-\dvc)(\divergence\avec)\vvech\right]}{(\uvechN+\uprime)} \\
		&= - \inner{\left[((\avech+\avec')\cdot\grad)\vvech + (1-\dvc)(\divergence(\avech+\avec'))\vvech\right]}{(\uvechN+\uprime)}
	\end{align}
	To retain the fine scale $\avec'$, one must be able to compute the divergence of $\avec'$. Isolating the term involving $\divergence\avec'$, we have
	\begin{align}
		-(1-\dvc)\inner{(\divergence\avec')\, \vvech}{(\uvechN+\uvec')} &= -(1-\dvc) \sum_{K \in \mesh} \inner{(\divergence\avec')\, \vvech}{(\uvechN+\uvec')}_K \\
		&= -(1-\dvc) \sum_{K \in \mesh} \inner{-(\divergence(\taum \resMom(\avech)))\, \vvech}{(\uvechN+\uvec')}_K.
	\end{align}
	Therefore, one needs to compute both $\grad\taum$ as well as $\divergence\resMom(\avech,p^{n-1})$. However, if one uses a stabilization parameter that is constant over an element (e.g., \cite{tezduyar2000finite}), then only the latter is needed.
\end{remark}

\subsection{Energy balance of the three advection forms}
\label{sec:energy-balance}

In this section, we provide a formal analysis of the energy stability of the semi-implicit method presented in \eqref{eq:oseen-vms} and \eqref{eq:oseen-vms-bdf}. Below, we first state our assumptions and other definitions. Then we present the general energy balance and the main stability conclusions for the three advection forms; the detailed derivations are given in Appendix~\ref{app:energy-derivation}.

\subsubsection{Time-continuous case}
\label{sec:energy-stability}
For the time-continuous case, we start from \eqref{eq:oseen-vms} and substitute the test function $\vvec$ by the trial function $\uvec$. The primary term of interest is the convection term, for which we have the following result.

\begin{proposition}[{Energy contribution of $\opAdvGen[\avec]{\dvc}$}]
	\label{thm:energy-identity}
	Let $\avec$ and $\uvec$ be sufficiently smooth with well-defined traces on
	$\partial\spDom$. Then
	\begin{align}
		\label{eq:energy-identity}
		\inner{\uvec}{\opAdvGen[\avec]{\dvc}\uvec}
		= \left(\dvc-\tfrac12\right)\inner{(\divergence\avec)\,\uvec}{\uvec}
		+ \tfrac12\int_{\partial\spDom}(\avec\cdot\normalvec)\,|\uvec|^2\,\ds .
	\end{align}
\end{proposition}

\begin{proof}
	Since the inner product in question is real, we have,
	\begin{align}
		\label{eq:energy-prop-1}
		\inner{\uvec}{\opAdvGen[\avec]{s}\uvec} = \inner{\opAdvGen[\avec]{s}\uvec}{\uvec}.
	\end{align}
	Furthermore, setting $\vvec=\uvec$ in \eqref{eq:adv-weaken-relation}, we have
	\begin{align}
		\label{eq:energy-prop-2}
		\inner{\uvec}{\opAdvGen[\avec]{s}\uvec}
		= -\inner{\opAdvGen[\avec]{1-s}\uvec}{\uvec} + \innerB{\uvec}{\normalvec\cdot(\uvec\avec^T)}.
	\end{align}
	Adding \eqref{eq:energy-prop-1} and \eqref{eq:energy-prop-2}, we have
	\begin{align*}
		2 \inner{\uvec}{\opAdvGen[\avec]{s}\uvec} &=  \inner{\opAdvGen[\avec]{s}\uvec}{\uvec} -  \inner{\opAdvGen[\avec]{1-s}\uvec}{\uvec} + \innerB{\uvec}{\normalvec\cdot(\uvec\avec^T)} \\
		&= (2\dvc-1)\inner{(\divergence\avec)\uvec}{\uvec} + \innerB{\uvec}{\normalvec\cdot(\uvec\avec^T)}.
	\end{align*}
	Dividing by 2 on both sides gives \eqref{eq:energy-identity}.
\end{proof}

\paragraph{Energy balance.} Now choosing the test pair $(\vvech,\qh)=(\uvech,\ph)$ in the time-continuous Oseen--VMS
formulation \eqref{eq:oseen-vms}, the pressure couplings cancel, the diffusion term
contributes $\visco\|\grad\uvech\|^2$, and the coarse-scale convection contribution is
exactly \eqref{eq:energy-identity}.
Then the time-continuous energy balance can be written as
\begin{multline}
	\label{eq:energy-balance-timecont}
	\frac{1}{2}\,\frac{d}{dt}\,\|\uvech\|^2
	+ \visco\,\|\grad\uvech\|^2
	+ \left(\dvc-\tfrac12\right)\inner{(\divergence\avech)\,\uvech}{\uvech} \\
	+ \tfrac12\int_{\Gamma_N}(\avech\cdot\normalvec)\,|\uvech|^2\,\ds
	+ \mathcal{S}_h
	\;=\; \inner{\fvec}{\uvech} + \inner{\hvec}{\uvech}_{\Gamma_N},
\end{multline}
where the $L^2(\spDom)$-norm is defined as
\begin{align}
	\norm{\wvec}^2 := \inner{\wvec}{\wvec},
\end{align}
and $\mathcal{S}_h$ collects every stabilization term of \eqref{eq:oseen-vms}, including those acting on $\fvec$. This
is an exact identity, and the stabilization terms $\mathcal{S}_h$ are general. In the next section, we analyze $\mathcal{S}_h$ in the discrete setting.

\subsubsection{Time-discrete case}
\label{sec:energy-discrete}
We now repeat the computation for the fully discrete scheme \eqref{eq:oseen-vms-bdf}. The stabilization parameters are those already defined in \eqref{eq:taum-tauc-shakib}; they are not redefined here. The analysis uses exactly two consequences of that definition, each obtained by discarding the remaining (non-negative) terms inside the bracket of \eqref{eq:taum-metric}:
\begin{align}
\label{eq:tau-bounds}
\taum \;\leq\; \frac{\dt}{2\beta_0},
\qquad\qquad
\taum \;\leq\; \left(\cinv\,\visco^2\,\mmat{G}\!:\!\mmat{G}\right)^{-\halfnice}
\;\leq\; \frac{h_K^2}{4\sqrt{d\,\cinv}\;\visco} ,
\end{align}
the last step using $\mmat{G}\!:\!\mmat{G} \geq 16\,d\,h_K^{-4}$ on a shape-regular mesh, with $h_K$ the element diameter and $\mmat{G}$ normalized as in Remark~\ref{rem:tau-uniform}. Nothing beyond positivity is used about $\tauc$.

\paragraph{Definitions and assumptions.}
Throughout, we take homogeneous Dirichlet data, $\uvec_D=\zerovec$ on $\Gamma_D$ and $\Gamma_N=\emptyset$, so that $\uvechN$ itself is an admissible test function in \eqref{eq:oseen-vms-bdf}. We write the BDF-$r$ difference operator of \eqref{eq:bdf-disc} as
\begin{align}
\label{eq:bdf-operator}
\bdfop\uvechN := \oneOver{\dt}\sum_{i=0}^{r}\beta_{-i}\,\uvech^{n-i},
\qquad\text{so that}\qquad
\sigma\uvechN - \gvecN \;=\; \bdfop\uvechN - \fvecN .
\end{align}
The momentum residual
\begin{align}
\label{eq:res-def}
\resMom^{(\dvc)}
\;:=\;
\bdfop\uvechN
+ \opAdvGen[\avech]{\dvc}\uvechN
- \visco\laplacian\uvechN
+ \grad\phN
- \fvecN
\end{align}
is thereby the residual of \eqref{eq:oseen} at time step $n$, and is the object appearing in the fine-scale closure \eqref{eq:oseen-timedisc-scale-residuals}, i.e.\ $\uprime=-\taum\resMom^{(\dvc)}$.

For a positive, element-wise-defined weight $\tau$, we use the broken weighted
inner product and norm
\begin{align}
  \inner{\vvec}{\wvec}_{\tau,h}
  &:= \sum_{K\in\mesh}\inner{\vvec}{\tau\wvec}_K,
  &
  \norm{\wvec}_{\tau,h}^2
  &:= \inner{\wvec}{\wvec}_{\tau,h}
   = \sum_{K\in\mesh}\norm{\tau^{1/2}\wvec}_K^2 .
  \label{eq:weighted-broken-norm}
\end{align}
We assume that the mesh $\mesh$ is shape-regular, such that the discrete element-wise inverse inequality
\begin{align}
\label{eq:inverse-estimate}
\norm{\laplacian\uvech}_K \;\leq\; \sqrt{\cinv}\,h_K^{-1}\,\norm{\grad\uvech}_K
\qquad\text{for all }\uvech\in\spaceVh
\end{align}
holds on the mesh, with explicit constants for inequalities of this form tabulated in \cite{harari1992what}. We further assume that (see Appendix~\ref{app:energy-crossterms})
\begin{align}
\label{eq:cond-inv}
\sqrt{\cinv/d} \;\leq\; \alpha \;<\; 8 .
\end{align}
\begin{proposition}[Discrete energy balance]
\label{thm:energy-discrete}
Assume $r\leq 2$. Then, the solution $\uvech$ of \eqref{eq:oseen-vms-bdf} satisfies, for all $\dvc\in \{0,\halfnice,1\}$,
\begin{multline}
  \label{eq:energy-general}
  \inner{\bdfop\uvechN}{\uvechN}
  + \left(1-\frac{\alpha}{8}\right)\visco\,\|\grad\uvechN\|^2
  + \norm{\divergence\uvechN}_{\tauc,h}^2
  + \frac{1}{8}\norm{\resMom^{(\dvc)}}_{\taum,h}^2
  + T_1^{(\dvc)}
  + T_2^{(\dvc)} \\
  \;\leq\;
  \frac{\dt}{2\beta_0}\norm{\bdfop\uvechN}^2
  + \inner{\fvecN}{\uvechN}
  + \frac{\dt}{\beta_0}\|\fvecN\|^2 ,
\end{multline}
where
\begin{align}
  \label{eq:T1-T2-def}
  T_1^{(\dvc)} := \left(\dvc-\tfrac12\right)
  \inner{(\divergence\avech)\,\uvechN}{\uvechN},
  \qquad
  T_2^{(\dvc)} := (1-2\dvc)
  \inner{(\divergence\avech)\,\uvechN}{\resMom^{(\dvc)}}_{\taum,h} .
\end{align}
\end{proposition}
\begin{proof}
See Appendix~\ref{app:energy-derivation}.
\end{proof}

\paragraph{Consequences for the three advection forms.}
Apart from $T_1^{(\dvc)}$ and $T_2^{(\dvc)}$, no term in \eqref{eq:energy-general} carries an explicit factor of $\dvc$, and every term on the left other than these two is non-negative. The energy behavior of the three advection forms therefore hinges entirely on $T_1^{(\dvc)}+T_2^{(\dvc)}$. Substituting $\dvc=\halfnice,0,1$ (Appendix~\ref{app:energy-specialization}) gives the following.
\begin{myitemize}
	\item \textbf{Skew-symmetric form} ($\dvc=\halfnice$): $T_1^{(1/2)}=T_2^{(1/2)}=0$ identically, so every term on the left-hand side of \eqref{eq:energy-general} is non-negative, for any mesh, any time step, and any $\|\divergence\avech\|_{L^\infty}$.
	\item \textbf{Convective form} ($\dvc=0$) \textbf{and divergence form} ($\dvc=1$): in both cases $T_1^{(\dvc)}$ and $T_2^{(\dvc)}$ are sign-indefinite, and no such conclusion is available. Absorbing $T_2^{(\dvc)}$ into the residual term of \eqref{eq:energy-general} leaves, in the worst case, the same deficit $-2\norm{(\divergence\avech)\uvechN}_{\taum,h}^2$ for \emph{both} values of $\dvc$ (Appendix~\ref{app:energy-specialization}).
\end{myitemize}

\subsection{Implementation details}
\label{sec:implementations}

The semi-implicit VMS formulation \eqref{eq:oseen-vms} requires only standard continuous Galerkin ingredients (Lagrange basis functions, element-wise residual evaluation, and element-wise stabilization parameters) and can be incorporated into existing FEM codebases. 
All numerical experiments in this paper are produced with an in-house MPI-based parallel finite element code written in C/C++. 
Mesh partitioning is performed with \textsc{ParMETIS}~\cite{karypis2003parmetis}. 
Parallel sparse linear algebra and solvers are provided by \textsc{PETSc}~\cite{balay2001petsc}. 

For the linear semi-implicit formulation, the algebraic system at each time step is solved with a Krylov method (KSP) using \texttt{BCGS} together with an additive Schwarz preconditioner (\texttt{ASM}). 
For the fully implicit nonlinear formulation, we use \textsc{PETSc} \texttt{SNES}; each Newton step is solved using the same Krylov and preconditioning options (\texttt{BCGS}+\texttt{ASM}). 
Additional details on the parallelization strategy and performance characteristics of the software can be found in~\cite{kodali2012computer,dyja2018parallel,khara2024solving}. Unless stated otherwise, identical solver tolerances are used when comparing linear and nonlinear variants, so that reported timing differences reflect the algorithmic cost rather than the stopping criteria.

\section{Numerical results}\label{sec:results}

This section evaluates the \emph{semi-implicit stabilized formulation} introduced in \secref{sec:formulation}. 
A central practical question is how the choice of advection form affects robustness and accuracy. 
Accordingly, we compare three linear semi-implicit methods corresponding to $s\in\{0,\tfrac12,1\}$. Across all methods, we report
\begin{enumerate}[itemsep=0pt, topsep=2pt]
	\item energy-decay behavior,
	\item temporal accuracy on a manufactured solution,
	\item benchmark accuracy on canonical laminar flows,
	\item performance on representative three-dimensional turbulent flows,
	\item computational cost relative to fully implicit nonlinear VMS.
\end{enumerate}

We also compare them against the fully implicit nonlinear VMS methods corresponding to $s\in\{0,1\}$. In particular, on a given spatial and temporal discretization, we treat the results from the nonlinear solver as baselines. Note that our primary goal in this section is to analyze and test the proposed linear semi-implicit method. The nonlinear results serve as reference.

\subsection{No-slip decay in a box}
\label{subsec:no_slip_decay_box}

\begin{figure}[htbp]
	\centering
	
	\begin{minipage}[t]{0.49\textwidth}
		\vspace{0pt}
		\centering
		\begin{tikzpicture}
			\begin{axis}[
				width=\linewidth,
				height=0.9\linewidth,
				xlabel={$t$},
				ylabel={Energy},
				title={$C=1.0$},
				xmin=-1,
				xmax=35,
				ymin=-0.2,
				ymax=2.5,
				grid=both,
				no marks,
				line width=1pt,
				legend columns=3,
				legend image post style={line width=1pt},
				legend style={
					at={(1.08,-0.22)},
					anchor=north,
					draw=black,
					fill=white,
				},
				]
				
				\addplot+[] table[col sep=comma, x=t, y=m1, each nth point=10]
				{data/dib/re_1e+03_C_1.0.txt};
				\addlegendentry{linear, $s=0$}
				
				\addplot+[] table[col sep=comma, x=t, y=m2, each nth point=10]
				{data/dib/re_1e+03_C_1.0.txt};
				\addlegendentry{linear, $s=1/2$}
				
				\addplot+[] table[col sep=comma, x=t, y=m3, each nth point=10]
				{data/dib/re_1e+03_C_1.0.txt};
				\addlegendentry{linear, $s=1$}
				
				\addplot+[dashed] table[col sep=comma, x=t, y=n1, each nth point=10]
				{data/dib/re_1e+03_C_1.0.txt};
				\addlegendentry{nonlinear, $s=0$}
				
				\addplot+[dashed] table[col sep=comma, x=t, y=n2, each nth point=10]
				{data/dib/re_1e+03_C_1.0.txt};
				\addlegendentry{nonlinear, $s=1$}
				
			\end{axis}
		\end{tikzpicture}
	\end{minipage}
	\hfill
	\begin{minipage}[t]{0.49\textwidth}
		\vspace{0pt}
		\centering
		\begin{tikzpicture}
			\begin{axis}[
				width=\linewidth,
				height=0.9\linewidth,
				xlabel={$t$},
				ylabel={Energy},
				title={$C=5.0$},
				xmin=-1,
				xmax=35,
				ymin=-0.2,
				ymax=2.5,
				grid=both,
				no marks,
				line width=1pt,
				legend image post style={line width=1pt},
				]
				
				\addplot+[] table[col sep=comma, x=t, y=m1, each nth point=10]
				{data/dib/re_1e+03_C_5.0.txt};
				
				\addplot+[] table[col sep=comma, x=t, y=m2, each nth point=10]
				{data/dib/re_1e+03_C_5.0.txt};
				
				\addplot+[] table[col sep=comma, x=t, y=m3, each nth point=2]
				{data/dib/re_1e+03_C_5.0.txt};
				
				\addplot+[dashed] table[col sep=comma, x=t, y=n1, each nth point=10]
				{data/dib/re_1e+03_C_5.0.txt};
				
				\addplot+[dashed] table[col sep=comma, x=t, y=n2, each nth point=10]
				{data/dib/re_1e+03_C_5.0.txt};
				
			\end{axis}
		\end{tikzpicture}
	\end{minipage}
	
	
	\caption{Energy decay for the no-slip decay-in-a-box problem at $\mathrm{Re}=10^3$ for two Courant numbers.}
	\label{fig:dib-energy-re1e3}
\end{figure}

We begin with the case of an incompressible flow inside a box \cite{zhang2016gepup,zhao2026amr_dirichlet_nse}. The domain is a unit square, i.e., $\Omega = [0,1]^2$, and has a no-slip boundary condition on the walls, i.e.,
\begin{align}
	\uvec=\zerovec \text{ on }\partial \Omega.
\end{align}
The flow is initialized by the divergence-free velocity field
\begin{align}
	\uvec_0(x,y)
	&=
	\begin{bmatrix}
		\pi \sin^2(\pi x)\sin(2\pi y) \\
		-\pi \sin(2\pi x)\sin^2(\pi y)
	\end{bmatrix}.
\end{align}
This initial condition is generated by the streamfunction
\begin{align}
	\psi_0(x,y)
	&=
	\sin^2(\pi x)\sin^2(\pi y),
\end{align}
through
\begin{align}
	\uvec_0
	&=
	\begin{bmatrix}
		\partial_y \psi_0 \\
		-\partial_x \psi_0
	\end{bmatrix}.
\end{align}
Hence, $\uvec_0$ is divergence-free by design. There is no body force, i.e., $\fvec=\zerovec$ in $\Omega$.

In the continuous setting (i.e., \eqref{eq:nse}), for $\fvec=\zerovec$, the kinetic energy
\begin{align}
	E(t)
	&=
	\frac{1}{2}\int_{\Omega} |\uvec(\xvec,t)|^2 \, d\xvec
\end{align}
satisfies the energy balance
\begin{align}
	\frac{dE}{dt}
	&=
	-\frac{1}{Re}
	\int_{\Omega} |\nabla \uvec(\xvec,t)|^2 \, d\xvec,
\end{align}
provided that the velocity is divergence-free and satisfies the homogeneous
Dirichlet boundary condition. Thus, the kinetic energy should decay
monotonically in time.

We solve this example with the linear semi-implicit scheme \eqref{eq:oseen-vms-bdf} for $\dvc \in \{0,\halfnice, 1\}$ for $Re=10^3$ on a $32\times 32$ mesh of axis-aligned regular linear quadrilateral elements, with a time step $\dt$ corresponding to the Courant number, $C$, given by
\begin{align}
	C := \frac{\norm{\avec}_{\infty} \dt}{h}.
\end{align} 
For this example, we take $\norm{\avec}_{\infty} = \norm{\uvec_0}_{\infty} = \pi$. 

\figref{fig:dib-energy-re1e3} shows the energy curves for these three linear methods with $C=1$ and $C=5$. Energy curves from the nonlinear schemes ($\dvc\in \{0,1\}$) are also shown for reference. For $C=1$, all schemes show a monotonic energy decay. Also, for $C=5$, all methods except $\dvc=1$ show monotonic decay. The linear scheme with $\dvc=1$ shows a non-monotonic behavior, where the energy rises and subsides multiple times in the time- interval $I=[0,35]$.

Therefore, this case serves as a simple counterexample which shows that $\dvc=1$ leads to a conditionally stable method. On the other hand, $\dvc\in \{0, \halfnice\}$ behave in a stable manner under the same conditions, closely mimicking the nonlinear methods.

\subsection{Convergence study}\label{sec:results-mms}

\begin{figure}[!htb]
	\centering
	\vspace{2.0em}
	\begin{minipage}[b]{.32\textwidth}
		\vspace{0pt}
		\centering
		\begin{tikzpicture}
			\begin{loglogaxis}[
				width=\linewidth,
				height=0.82\linewidth,
				xlabel=$\nicefrac{dt}{T}$,
				legend style={
					at={(0,1)},
					anchor=north west,
					legend columns=1,
					font=\tiny,
					draw=black,
					fill=white,
				},
				x tick label style={rotate=0,anchor=north},
				xtick={1/256, 1/128, 1/64, 1/32, 1/16},
				xticklabels={$2^{-8}$,$2^{-7}$,$2^{-6}$,$2^{-5}$,$2^{-4}$},
				]
				\addplot+ table[
				x expr={(1.0 / \thisrow{nsteps})},
				y expr={sqrt((\thisrow{err_x})^2 + (\thisrow{err_y})^2)},
				col sep=comma
				]{data/mmsHarmonic/linear_s_0.0_re_1e+03.txt};
				
				\addplot+[dashed] table[
				x expr={(1.0 / \thisrow{nsteps})},
				y expr={sqrt((\thisrow{err_x})^2 + (\thisrow{err_y})^2)},
				col sep=comma
				]{data/mmsHarmonic/linear_s_0.0_re_1e+06.txt};
				
				\logLogSlopeTriangle{0.8}{0.2}{0.2}{2}{blue};
				
				\legend{
					$Re = 10^3$,
					$Re = 10^6$
				}
			\end{loglogaxis}
		\end{tikzpicture}
		\subcaption{Linear, $\dvc=0$}
		\label{fig:s-0.0-mmsHarmonic}
	\end{minipage}
	\hfill
	\begin{minipage}[b]{.32\textwidth}
		\vspace{0pt}
		\centering
		\begin{tikzpicture}
			\begin{loglogaxis}[
				width=\linewidth,
				height=0.82\linewidth,
				xlabel=$\nicefrac{dt}{T}$,
				x tick label style={rotate=0,anchor=north},
				xtick={1/256, 1/128, 1/64, 1/32, 1/16},
				xticklabels={$2^{-8}$,$2^{-7}$,$2^{-6}$,$2^{-5}$,$2^{-4}$},
				]
				\addplot+ table[
				x expr={(1.0 / \thisrow{nsteps})},
				y expr={sqrt((\thisrow{err_x})^2 + (\thisrow{err_y})^2)},
				col sep=comma
				]{data/mmsHarmonic/linear_s_0.5_re_1e+03.txt};
				
				\addplot+[dashed] table[
				x expr={(1.0 / \thisrow{nsteps})},
				y expr={sqrt((\thisrow{err_x})^2 + (\thisrow{err_y})^2)},
				col sep=comma
				]{data/mmsHarmonic/linear_s_0.5_re_1e+06.txt};
				
				\logLogSlopeTriangle{0.8}{0.2}{0.2}{2}{blue};
			\end{loglogaxis}
		\end{tikzpicture}
		\subcaption{Linear, $\dvc=\halfnice$}
		\label{fig:s-0.5-mmsHarmonic}
	\end{minipage}
	\hfill
	\begin{minipage}[b]{.32\textwidth}
		\vspace{0pt}
		\centering
		\begin{tikzpicture}
			\begin{loglogaxis}[
				width=\linewidth,
				height=0.82\linewidth,
				xlabel=$\nicefrac{dt}{T}$,
				x tick label style={rotate=0,anchor=north},
				xtick={1/256, 1/128, 1/64, 1/32, 1/16},
				xticklabels={$2^{-8}$,$2^{-7}$,$2^{-6}$,$2^{-5}$,$2^{-4}$},
				]
				\addplot+ table[
				x expr={(1.0 / \thisrow{nsteps})},
				y expr={sqrt((\thisrow{err_x})^2 + (\thisrow{err_y})^2)},
				col sep=comma
				]{data/mmsHarmonic/linear_s_1.0_re_1e+03.txt};
				
				\addplot+[dashed] table[
				x expr={(1.0 / \thisrow{nsteps})},
				y expr={sqrt((\thisrow{err_x})^2 + (\thisrow{err_y})^2)},
				col sep=comma
				]{data/mmsHarmonic/linear_s_1.0_re_1e+06.txt};
				
				\logLogSlopeTriangle{0.8}{0.2}{0.2}{2}{blue};
			\end{loglogaxis}
		\end{tikzpicture}
		\subcaption{Linear, $\dvc=1$}
		\label{fig:s-1.0-mmsHarmonic}
	\end{minipage}
	
	\caption{Convergence of the velocity error magnitude with the semi-implicit methods ($T = 2.25$).}
	\label{fig:mms-conv-linear}
\end{figure}

\begin{figure}[!htb]
	\centering
	\vspace{2.0em}
	\begin{minipage}[b]{.32\textwidth}
		\vspace{0pt}
		\centering
		\begin{tikzpicture}
			\begin{loglogaxis}[
				width=\linewidth,
				height=0.82\linewidth,
				xlabel=$\nicefrac{dt}{T}$,
				legend style={
					at={(0,1)},
					anchor=north west,
					legend columns=1,
					font=\tiny,
					draw=black,
					fill=white,
				},
				x tick label style={rotate=0,anchor=north},
				xtick={1/256, 1/128, 1/64, 1/32, 1/16},
				xticklabels={$2^{-8}$,$2^{-7}$,$2^{-6}$,$2^{-5}$,$2^{-4}$},
				]
				\addplot+ table[
				x expr={(1.0 / \thisrow{nsteps})},
				y expr={sqrt((\thisrow{err_x})^2 + (\thisrow{err_y})^2)},
				col sep=comma
				]{data/mmsHarmonic/nonlinear_s_0.0_re_1e+03.txt};
				
				\addplot+[dashed] table[
				x expr={(1.0 / \thisrow{nsteps})},
				y expr={sqrt((\thisrow{err_x})^2 + (\thisrow{err_y})^2)},
				col sep=comma
				]{data/mmsHarmonic/nonlinear_s_0.0_re_1e+06.txt};
				
				\logLogSlopeTriangle{0.8}{0.2}{0.2}{2}{blue};
				
				\legend{
					$Re = 10^3$,
					$Re = 10^6$
				}
			\end{loglogaxis}
		\end{tikzpicture}
		\subcaption{Nonlinear, $\dvc=0$}
		\label{fig:nonlinear-s-0.0-mmsHarmonic}
	\end{minipage}
	\hspace{0.06\textwidth}
	\begin{minipage}[b]{.32\textwidth}
		\vspace{0pt}
		\centering
		\begin{tikzpicture}
			\begin{loglogaxis}[
				width=\linewidth,
				height=0.82\linewidth,
				xlabel=$\nicefrac{dt}{T}$,
				x tick label style={rotate=0,anchor=north},
				xtick={1/256, 1/128, 1/64, 1/32, 1/16},
				xticklabels={$2^{-8}$,$2^{-7}$,$2^{-6}$,$2^{-5}$,$2^{-4}$},
				]
				\addplot+ table[
				x expr={(1.0 / \thisrow{nsteps})},
				y expr={sqrt((\thisrow{err_x})^2 + (\thisrow{err_y})^2)},
				col sep=comma
				]{data/mmsHarmonic/nonlinear_s_1.0_re_1e+03.txt};
				
				\addplot+[dashed] table[
				x expr={(1.0 / \thisrow{nsteps})},
				y expr={sqrt((\thisrow{err_x})^2 + (\thisrow{err_y})^2)},
				col sep=comma
				]{data/mmsHarmonic/nonlinear_s_1.0_re_1e+06.txt};
				
				\logLogSlopeTriangle{0.8}{0.2}{0.2}{2}{blue};
			\end{loglogaxis}
		\end{tikzpicture}
		\subcaption{Nonlinear, $\dvc=1$}
		\label{fig:nonlinear-s-1.0-mmsHarmonic}
	\end{minipage}
	
	\caption{Convergence of the velocity error magnitude with the nonlinear methods ($T=2.25$).}
	\label{fig:mms-conv-nonlinear}
\end{figure}

We next test the convergence of the method proposed in \eqref{eq:oseen-vms} by examining how the errors behave with respect to the time-step size $\dt$. We take the exact solution $ (\uvec, p) $ to be

\begin{subequations}
\label{eq:mms-1}
\begin{align}
\uvec &= 
\begin{bmatrix}
\sin(\pi x) \cos(\pi y) \sin(2 \pi t) \\
-\cos(\pi x) \sin(\pi y) \sin(2 \pi t) \\
\end{bmatrix},\\
	p &= \sin(\pi x) \sin(\pi y) \cos(2 \pi t)
\end{align}
\end{subequations}
in a unit square domain, i.e., $\spDom = [0,1]^2$ and time interval $I_T=[0,T]$. We set $T=2.25$, since the velocities reach a peak at $t=n + \tfrac{1}{4}$ for $n\in\{0,1,2,\ldots\}$. The velocity $\uvec$ is chosen such that it satisfies $ \grad\cdot\uvec = 0 $ exactly pointwise. We obtain the symbolic / analytical expression for the forcing $ \fvec $ by substituting the exact solutions $(\uvec, p)$ in \eqref{eq:nse}. We also analytically obtain the initial condition $\uvec_0$ by setting $t=0$ in \eqref{eq:mms-1}, and the boundary condition $\uvec_d$ on all sides by setting $\xvec = (x,y)$ appropriately.

Once we have the expressions for $\fvec$, $\uvec_0$ and $\uvec_d$, we solve \eqref{eq:nse} using the proposed numerical method \eqref{eq:oseen-vms-bdf}, and obtain the approximate solutions $(\uvech, \ph)$. Finally, we measure the error in $(\uvech, \ph)$ with respect to the exact solution $(\uvec, p)$ using
\begin{align}
\norm{e(t)} = \left[ \int_{\spDom} \big| \uvech(t) - \uvec(t) \big|^2\ \mbox{d}\xvec\right]^{\halfnice}.
\end{align}

\figref{fig:mms-conv-linear} shows the convergence of $\norm{e(T)}$ as the step size $\dt$ is varied from $\dt=T/16$ to $\dt=T/256$, for $\dvc = 0,\ 0.5,\ 1$, each at two different Reynolds numbers 
$Re = 10^3$ and $10^6$. The time discretization is BDF2, so we expect an order of convergence of 2. This order of convergence is achieved for $\dvc=0$ and $\dvc=0.5$. However, $\dvc=1$ exhibits an unstable behavior, and errors do not necessarily decrease when the time step is reduced. The same convergence study for the nonlinear methods for both $\dvc=0,1$ is presented in \figref{fig:mms-conv-nonlinear} for reference. Both formulations show an order of convergence of 2 as $\dt$ is reduced.

\subsection{Lid-driven cavity (2D)} \label{sec:ldc2d}

\figref{fig:schematic-ldc2d} shows a schematic of the lid-driven cavity problem in 2D. The computational domain is $\spDom = [0,1]^2$. The boundary conditions are static (do not change with time) and are given by
\begin{subequations}
\begin{align}
\uvec &= \zerovec,\ y=0,\\
\uvec &= [1,0]^T,\ y=1,\\
\uvec &= \zerovec\ x\in\{0,\ 1\}.
\end{align}
\end{subequations}
The boundary conditions comprise a ``lid'' at the top moving with horizontal velocity $=1$, and no-slip walls on the rest of the three sides. Since the pressure is determined only up to an additive constant, we set (see \figref{fig:schematic-ldc2d})
\begin{align}
    p=0 \ \text{at} \ \xvec=\zerovec.
\end{align}
The forcing $\fvec = \zerovec$, and the initial condition is given by a zero function
\begin{align}
\uvec_0 = \zerovec.
\end{align}

\begin{figure}[!htb]
\centering
\begin{tikzpicture}[scale=4]

\draw[thick] (0,0) -- (1,0) -- (1,1) -- (0,1) -- cycle;

\node[align=center] at (0.5,-0.15) {$y=0,$\\$\uvec = \zerovec$}; 
\node[align=center] at (0.5,1.15) {$y=1,$\\$\uvec = [1,0]$};
\node[align=center] at (-0.25,0.5) {$x=0$,\\$\uvec = \zerovec$}; 
\node[align=center] at (1.25,0.5) {$x=1$,\\$\uvec = \zerovec$};

\draw[dashed, very thick, blue!90] (0.5,0) -- (0.5,1);
\draw[dashed, very thick, blue!90] (0,0.5) -- (1,0.5);
\node[blue!90] at (0.55,0.15) {$A$};
\node[blue!90] at (0.15,0.55) {$B$};

\node at (-1,0.5) {$\Omega=[0,1]^2$};

\draw[dotted,very thick] (0,0) circle(0.07);

\fill (0,0) circle(0.015);

\node[align=center] at (-0.4,-0.25) (pzero) {$\xvec=\zerovec$,\\$p=0$};
\draw[->,thick,>=stealth,line width=1pt] (pzero) -- (0,0);

\def\axOx{-1}
\def\axOy{0}
\def\axL{0.2}
\draw[very thick,->,>=stealth] (\axOx,\axOy) -- (\axOx+\axL,\axOy);
\draw[very thick,->,>=stealth] (\axOx,\axOy) -- (\axOx,\axOy+\axL);
\node at (\axOx+\axL,\axOy) [right] {$x$};
\node at (\axOx,\axOy+\axL) [left] {$y$};

\end{tikzpicture}
\caption{Boundary conditions for the lid-driven cavity problem. The blue dashed lines are the midlines where velocities will be evaluated and plotted (see \figref{fig:ldc-re-all-panel-1})}
\label{fig:schematic-ldc2d}
\end{figure}
\input{supporting_tex/ldc_figures_split.tex}

We solve this problem for a series of Reynolds numbers $ Re = $ 100, 1000, 5000 and 10000. The benchmark results for this problem generally refer to the steady-state solution. Suppose $ \partial_t \uvec = \zerovec,\ \partial_t p = 0 $ for some $ t \geq t_s $. Then the final time horizon $T$ is chosen such that $T>t_s$. Usually $ t_s $ increases with an increase in the Reynolds number. Here, we choose $ T = 5\times 10^3 $ for all cases. The mesh is a $128\times 128$ linear quadrilateral mesh, where the spacing between the elements varies according to a cosine curve, such that the elements near the boundaries are more refined than those near the center. We use $\dt=1$ for $Re < 5000$, $\dt=0.5$ for $ 5000 \leq Re \leq 7000$ and $\dt=0.1$ for $Re > 7000$.

\figref{fig:ldc-re-all-panel-1} plots the standard midline velocity profiles for all the considered Reynolds numbers from the linear scheme with $\dvc = 0,\ \halfnice$. Results from the nonlinear method with $\dvc=0,\ 1$ are also presented for reference. \figref{fig:ldc-re-ux} shows the $x$-component of the velocity ($u_x$) on the $x=0.5$ vertical line (line `A' in \figref{fig:schematic-ldc2d}). And similarly, \figref{fig:ldc-re-uy} plots the $y$-component of the velocity ($u_y$) on the $y=0.5$ horizontal line (line `B' in \figref{fig:schematic-ldc2d}).

The linear scheme with $\dvc=1$ is once again an exception. While this scheme yields comparable results for lower $Re$ values, it produces spurious oscillations for higher ones. We present the line-cuts for $Re=5000$ in \figref{fig:ldc-re5000-linear-s1}. On the same $128\times 128$ mesh, there are oscillations near the boundaries. However, when the mesh is refined to a $200\times 200$ grid (similarly graded as above), the oscillations disappear.

Therefore, we see that the linear methods ($\dvc=0,\halfnice$) are in good agreement with the reference values as well as the nonlinear methods ($\dvc=0,1$) across the full range of Reynolds numbers on the resolution used here. The exception is the linear method with $s=1$ which requires a finer mesh on higher Reynolds numbers.

\subsection{Flow past a circular cylinder (2D)}

\begin{figure}[!htb]
\centering
\begin{tikzpicture}[scale=4]

\draw[thick] (0,0) rectangle (3,1);

\node at (1.5,-0.15)[align=center] {$y=0$,\\$\uvec=\zerovec$}; 
\node at (1.5,1.15)[align=center] {$y=H$,\\$\uvec=\zerovec$};
\node at (-0.25,0.5) [rotate=90, align=center] {$x=0$,\\$\uvec=[\gamma(y),0]^T$}; 
\node at (3.2,0.5) [rotate=90, align=center] {$x=L$,\\$\normalvec\cdot(-p \mmat{I} + \visco \grad \uvec) = 0$};

\draw[thick] (1.0,0.5) circle(0.2);

\fill (1.0,0.5) circle(0.015);

\node at (1.0,0.18)[align=center] {cylinder,\\$\uvec=\zerovec$};

\draw[thick,->,>=stealth] (1.0,0.5) -- (1.0+0.141,0.5+0.141);
\draw[thick,->,>=stealth] (1.0,0.5) -- (1.0-0.141,0.5-0.141);

\node at (1.0+0.25,0.5+0.24) {$d=1$};

\def\xvert{0.15}

\draw[very thick,blue!90] (\xvert,0) -- (\xvert,1);

\fill[blue!50,opacity=0.5]
(\xvert,0)
\foreach \y in {0,0.01,...,1} {
-- ({\xvert + 0.25*4*\y*(1-\y)},\y)
}
-- (\xvert,1) -- (\xvert,0);

\draw[very thick,blue!90]
plot[domain=0:1,samples=100] 
({\xvert + 0.25*4*\x*(1-\x)},\x);

\foreach \y in {0.08,0.16,...,1} {
\pgfmathsetmacro\xparabola{\xvert + 0.25*4*\y*(1-\y)}
\draw[very thick,->,>=stealth,blue!90] (\xvert,\y) -- (\xparabola,\y);
}
\node at (0.45,0.15) [very thick,blue!90]{$\gamma(y)$};

\def\axO{-0.25}
\def\axL{0.2}
\draw[very thick,->,>=stealth] (\axO,\axO) -- (\axO+\axL,\axO);
\draw[very thick,->,>=stealth] (\axO,\axO) -- (\axO,\axO+\axL);
\node at (\axO+\axL,\axO) [right] {$x$};
\node at (\axO,\axO+\axL) [left] {$y$};

\end{tikzpicture}
\caption{Schematic diagram of the flow past a circular cylinder in 2D. The inlet velocityprofile $\gamma(y)$ is shown in blue curves and text.}
\label{fig:fpc2d-schematic}
\end{figure}

\begin{figure}[!htb]
		\centering
		\begin{minipage}{0.4\textwidth}
		\centering
		\begin{tikzpicture}
		\begin{axis}[width=0.99\linewidth,
		scaled y ticks=true,
		xlabel={$Re$},
		ylabel={$C_d$},
		legend style={at={(0.5,-0.25)},anchor=north, nodes={scale=0.65, transform shape}}, 
		legend pos=north east, 
		ymax=1.7,
		restrict x to domain=60:300,
		unbounded coords=discard,
		]
		\addplot+[solid, semithick, mark=o, mark options={fill=white}] table [x expr=\thisrowno{0},y expr=\thisrowno{1},col sep=comma] {data/fpcKanaris/trimesh/forces_linear_s_0.0.txt}; \addlegendentry{linear, $s=0$}
		\addplot+[solid, semithick, mark=square, mark options={fill=white}] table [x expr=\thisrowno{0},y expr=\thisrowno{1},col sep=comma] {data/fpcKanaris/trimesh/forces_linear_s_0.5.txt}; \addlegendentry{linear, $s=\halfnice$}
		\addplot+[solid, semithick, mark=triangle, mark options={fill=white}] table [x expr=\thisrowno{0},y expr=\thisrowno{1},col sep=comma] {data/fpcKanaris/trimesh/forces_linear_s_1.0.txt}; \addlegendentry{linear, $s=1$}
		\addplot+[solid, semithick, mark=diamond, mark options={fill=white}] table [x expr=\thisrowno{0},y expr=\thisrowno{1},col sep=comma] {data/fpcKanaris/trimesh/forces_nonlinear_s_0.0.txt}; \addlegendentry{nonlinear, $s=0$}
		\addplot+[solid, semithick, mark=pentagon, mark options={fill=white}] table [x expr=\thisrowno{0},y expr=\thisrowno{1},col sep=comma] {data/fpcKanaris/trimesh/forces_nonlinear_s_1.0.txt}; \addlegendentry{nonlinear, $s=1$}
		\addplot[only marks, mark=diamond*, mark options={fill=black, draw=black, scale=1.2}, color=black] table [x expr=\thisrowno{0},y expr=\thisrowno{1},col sep=comma] {data/fpcKanaris/reference/singha-drag-data.txt}; \addlegendentry{Singha et al. \ \protect\cite{singha2010flow}}
		\addplot[only marks, mark=square*, mark options={fill=black, draw=black, scale=1.0}, color=black] table [x expr=\thisrowno{1},y expr=\thisrowno{2},col sep=comma] {data/fpcKanaris/reference/zovatto-drag-data.txt}; \addlegendentry{Zovatto et al. \cite{zovatto2001flow}}
		\addplot[only marks, mark=asterisk, color=black] table [x expr=\thisrowno{0},y expr=\thisrowno{1},col sep=comma] {data/fpcKanaris/reference/kanaris-drag-data.txt}; \addlegendentry{Kanaris et al. \cite{kanaris2011three}}
		\end{axis}
		\end{tikzpicture}
		\subcaption{Drag coefficient}
		\label{fig:drag}
		\end{minipage}
		\hspace{5em}
		\begin{minipage}{0.4\textwidth}
		\centering
		\begin{tikzpicture}
		\begin{axis}[width=0.99\linewidth,
		scaled y ticks=true,
		xlabel={$Re$},
		ylabel={$St$},
		legend style={at={(0.5,-0.25)},anchor=north, nodes={scale=0.65, transform shape}}, 
		legend pos=south east, 
		ymin=0.14,ymax=0.21,
		restrict x to domain=60:300,
		unbounded coords=discard,
		]
		\addplot+[solid, semithick, mark=o, mark options={fill=white}] table [x expr=\thisrowno{0},y expr=\thisrowno{3},col sep=comma] {data/fpcKanaris/trimesh/forces_linear_s_0.0.txt}; \addlegendentry{linear, $s=0$}
		\addplot+[solid, semithick, mark=square, mark options={fill=white}] table [x expr=\thisrowno{0},y expr=\thisrowno{3},col sep=comma] {data/fpcKanaris/trimesh/forces_linear_s_0.5.txt}; \addlegendentry{linear, $s=\halfnice$}
		\addplot+[solid, semithick, mark=triangle, mark options={fill=white}] table [x expr=\thisrowno{0},y expr=\thisrowno{3},col sep=comma] {data/fpcKanaris/trimesh/forces_linear_s_1.0.txt}; \addlegendentry{linear, $s=1$}
		\addplot+[solid, semithick, mark=diamond, mark options={fill=white}] table [x expr=\thisrowno{0},y expr=\thisrowno{3},col sep=comma] {data/fpcKanaris/trimesh/forces_nonlinear_s_0.0.txt}; \addlegendentry{nonlinear, $s=0$}
		\addplot+[solid, semithick, mark=pentagon, mark options={fill=white}] table [x expr=\thisrowno{0},y expr=\thisrowno{3},col sep=comma] {data/fpcKanaris/trimesh/forces_nonlinear_s_1.0.txt}; \addlegendentry{nonlinear, $s=1$}
		\addplot[only marks, mark=square*, mark options={fill=black, draw=black, scale=1.0}, color=black] table [x expr=\thisrowno{1},y expr=\thisrowno{3},col sep=comma] {data/fpcKanaris/reference/zovatto-drag-data.txt}; \addlegendentry{Zovatto et al. \cite{zovatto2001flow}}
		\addplot[only marks, mark=asterisk, color=black] table [x expr=\thisrowno{0},y expr=\thisrowno{1},col sep=comma] {data/fpcKanaris/reference/kanaris-st-data.txt}; \addlegendentry{Kanaris et al. \cite{kanaris2011three}}
		\end{axis}
		\end{tikzpicture}
		\subcaption{Strouhal number}
		\label{fig:st}
		\end{minipage}
\caption{{Results for flow past a circular cylinder in two spatial dimensions:} (a) Drag coefficient $ C_d $, (b) Strouhal number $ St $}
\label{fig:fpc2d-validation}
\end{figure}

Next, we test our method on the problem of two-dimensional flow past a circular 
cylinder. The setup is taken from \cite{schafer1996benchmark,kanaris2011three,zovatto2001flow}. The schematic of the problem is shown in
\figref{fig:fpc2d-schematic}. The spatial domain is $ \spDom = ((I_x\times I_y) \backslash C) $, where $ I_x = [0,L],\ I_y = [0,H] $ and $ C $ denotes a circle of diameter $ d=1 $ with center at $ (x_c,y_c) $. We set $L=20$, $H=5$, $x_c=3$ and $y_c=2.5$.
The boundary and initial conditions are given as:
\begin{subequations}
	\label{eq:fpc-boundary-conditions}
	\begin{align}
	\uvec &= [\gamma(y), 0]^T\ \text{on}\ x=0,\\
\uvec &= [0,0]^T\ \text{on}\ y = 0,\ H,\\
	\uvec &= [0,0]^T\ \text{on cylinder}\ \Gamma_C,\\
\normalvec\cdot(-p \mmat{I} + \visco \grad \uvec) &= 0\ \text{on}\ x=L,
	\end{align}
\end{subequations}
where the inlet flow profile $ \gamma(y) $ is given by
\begin{align}\label{eq:fpc-inlet-vel}
	\gamma(y) = 4 u_c y (H-y) / H^2,
\end{align}
which is a parabolic profile with $ \gamma(0) = \gamma(H)=0 $ and $ \gamma_m := \gamma(\nicefrac{H}{2}) = u_c $. The average inlet velocity is given by $ \bar{\gamma} = \nicefrac{2}{3} u_c $. The Reynolds number for this problem is defined as $ Re = \frac{U_{ref} d}{\nu}$. $ U_{ref} = \gamma_m = u_c $. The force on the cylinder in the direction $ \wvec $ is expressed as
\begin{align}
F_w = \int_{\Gamma_C} \left[ \left( -p \mmat{I} + \visco [\grad \uvec + \grad \uvec^T] \right) \cdot \normalvec \right]\cdot \wvec \ dS.
\end{align}
We set $ \wvec = (1,0) $ to extract the drag force, and $ \wvec = (0,1) $ to extract the lift force. The drag coefficient $C_d$ and the lift coefficient $C_l$ are then calculated as
\begin{align}
C_d &= \frac{2 F_{\mathrm{drag}}}{AU^2} = 2 F_{\mathrm{drag}}, \\
C_l &= \frac{2 F_{\mathrm{lift}}}{AU^2} = 2 F_{\mathrm{lift}}
\end{align}
where $A=1$ and $U=1$ are the reference area and speed respectively.

After a sufficient time has passed, the solution of this example is known to reach either a steady state or a periodic state depending on whether the Reynolds number is below or above a critical value. When the solution achieves a periodic state, so do the drag and lift forces calculated above. Suppose $T_p$ is the time period of the lift coefficient $C_l$ obtained from the numerical simulation. Then the corresponding Strouhal number is calculated as
\begin{align}
St = \frac{L_r}{U_r T_p} = \frac{1}{T_p}
\end{align}
where $L_r=1$ and $U_r=1$ are the reference length and the speed respectively.

We solve this problem numerically for various $Re \in \{60, 100, 150, 200, 250\}$ on a relatively coarse mesh of 5600 linear quadrilaterals. We use $\dt=0.1$ for $Re \in (50,100]$, $\dt=0.05$ for $Re \in (100,200]$ and $\dt=0.03$ for $Re \in (200,250]$. We compute the drag coefficient $C_d$ and the Strouhal number $St$ for each case. \figref{fig:fpc2d-validation} shows a plot of these $C_d$ and $St$ values (along with reference values from the literature).

Note that, in this example, all the linear methods ($\dvc=0,\halfnice,1$) perform approximately the same. The unstable behavior encountered for $\dvc=1$ when all sides have Dirichlet conditions is absent in this case.

\subsection{Turbulent channel flow (3D)}

\begin{figure}[!htb]
\centering

\begin{tikzpicture}[scale=0.4, line cap=round, line join=round, >=stealth]

  \def\xshift{9}     
  \def\L{7*pi}       
  \def\H{6}
  \def\D{2*pi}

  \coordinate (A) at (\xshift,0);
  \coordinate (B) at (\xshift+\L,0);
  \coordinate (C) at (\xshift+\L,\H);
  \coordinate (Dpt) at (\xshift,\H);

  \coordinate (E) at ($(A)+(\D*0.5, \D*0.5)$);
  \coordinate (F) at ($(B)+(\D*0.5, \D*0.5)$);
  \coordinate (G) at ($(C)+(\D*0.5, \D*0.5)$);
  \coordinate (Hpt) at ($(Dpt)+(\D*0.5, \D*0.5)$);

  \draw[black,thick] (A)--(B)--(C)--(Dpt)--cycle;

  \draw[black,dashed] (E)--(F);
  \draw[black,thick]  (F)--(G)--(Hpt);
  \draw[black,dashed] (Hpt)--(E);

  \draw[black,thick] (B)--(F) (C)--(G) (Dpt)--(Hpt);
  \draw[black,dashed] (A)--(E); 

  \draw[->,thick] ($(A)+(0.4*\L,0.3*\H)$) -- ($(A)+(0.6*\L,0.3*\H)$)
    node[midway,above,fill=white,inner sep=1pt] {flow};

  \coordinate (O) at (\xshift-7,1.2); 
  \draw[->,thick] (O) -- ++(3,0) node[below right] {$x$};
  \draw[->,thick] (O) -- ++(0,3) node[above left] {$y$};
  \draw[->,thick] (O) -- ++(-1.6,-1.6) node[below left] {$z$}; 

  \draw[black,<->] (\xshift,-0.8) -- (\xshift+\L,-0.8)
    node[midway,fill=white,inner sep=2pt] {$ L$};

  \draw[black,<->] (\xshift-1.5,0) -- (\xshift-1.5,\H)
    node[midway,fill=white,inner sep=2pt] {$ H$};

  \draw[black,<->] (\xshift+\L+2.0,0) -- ++(\D*0.5,\D*0.5)
    node[midway,fill=white,inner sep=2pt] {$ D$};

\end{tikzpicture}

\caption{Turbulent channel geometry. Periodic boundary conditions are applied on the streamwise ($x$) and the spanwise ($z$) directions, whereas no-slip conditions are applied on the walls ($y$-boundaries).}
~\label{fig:channel_geometry}
\end{figure}

Starting from this section, we extend our numerical tests to 3D cases. For ease of representation, in this section, we will use $\xvec = [x,y,z]^T$ and $\uvec = [u_x,u_y,u_z]^T$. The first test case is the turbulent flow between two stationary parallel plates~\cite{john2016finite, moser1999direct, gravemeier2006scale} (the geometry is shown in \figref{fig:channel_geometry}). The domain is $\Omega = [0,L] \times [0,H] \times [0,D]$, where $L = 2\pi$, $H = 2$, and $D = \pi$. No-slip conditions are applied on the walls at $y=0$ and $y=H$, and periodic conditions are applied in the streamwise ($x$) and spanwise ($z$) directions. The flow is driven by a constant body force $\fvec = [1,0,0]^T$ in \eqref{eq:nse}.

The governing equations \eqref{eq:nse} are written in terms of the kinematic viscosity $\nu$, so the friction Reynolds number in our simulation is set through $\nu$. Taking the channel half-width $\delta = H/2 = 1$ as the characteristic length, the body force gives the friction velocity $u_\tau = \sqrt{f_x\,\delta} = 1$, so that $Re_\tau = u_\tau \delta / \nu = 1/\nu$. We set $\nu = 1/395$, corresponding to $Re_\tau = 395$.

The initial velocity field is constructed by perturbing the mean velocity profile obtained from the DNS data of~\cite{moserMKMdatabase}; random fluctuations with magnitude up to $10\%$ of the bulk velocity are added to all three velocity components, following~\cite{john2016finite}. The domain is discretized with a $64\times 64\times 64$ mesh that is uniform in the streamwise and spanwise directions and stretched in the wall-normal direction with clustering near the walls. We use $\dt = 10^{-3}$.

After the flow reaches a statistically steady state, the mean streamwise velocity $\langle u_x \rangle(y)$ is obtained by averaging $u_x$ in time and over the streamwise and spanwise directions, along which the flow is homogeneous. The fluctuation is then defined as $u_x' = u_x - \langle u_x \rangle$. Here, the superscript prime denotes the turbulent fluctuation and not the fine-scale model velocity mentioned in \secref{sec:stabilized-formulation-nse}. \figref{fig:cf3d-re395-plots} shows the mean velocity profile and \figref{fig:cf3d-re395-fluctuations} shows the root-mean-square (rms) velocity fluctuations obtained from the linear methods ($\dvc=0,\halfnice,1$), compared with the DNS data of \cite{moser1999direct}. Results from the nonlinear methods ($\dvc=0,1$) are also presented for reference, along with the residual-based VMS results of Bazilevs et al.~\cite{bazilevs2007variational} computed on a $64^3$ trilinear discretization. In the plots, $y^+$ and $\langle u_x \rangle^+$ denote wall-normal distance and mean streamwise velocity expressed in viscous wall units, following the standard definitions used in \cite{moser1999direct}.

We observe reasonable agreement with the DNS data across all methods. The mean velocity profile matches the DNS through the viscous sublayer and the buffer layer, and is over-predicted in the logarithmic region. The velocity fluctuations reproduce the shape of the DNS profiles and the correct peak locations, with the streamwise component over-predicted and the wall-normal component slightly under-predicted. These deviations are expected, since the reference data of \cite{moser1999direct} are fully resolved direct numerical simulations, whereas the present computations use a $64\times64\times64$ mesh on which the near-wall scales are not resolved and their effect enters only through the VMS closure \cite{Hughes01b, Akkerman08a, Bazilevs10d}. The linear and nonlinear methods follow the same trends, and the three linear forms are nearly indistinguishable from one another, indicating that the deviations are governed by the mesh resolution rather than by the linearization. Furthermore, our results closely align with the VMS results of Bazilevs et al.~\cite{bazilevs2007variational} obtained at the same resolution.

\begin{figure}[!htb]
\centering
\begin{tikzpicture}
\begin{axis}[
    width=0.95\textwidth,
    height=0.42\textwidth,
    xlabel={$y^+$},
    ylabel={$\langle u_x \rangle^+$},
    xmin=0.1,
    xmax=450,
    ymin=0,
    ymax=25,
    xmode=log,
    grid=both,
    grid style={dotted},
    line width=1pt,
    mark size=3pt,
    every mark/.append style={solid, line width=0.7pt},
    tick label style={font=\normalsize},
    label style={font=\normalsize},
    legend style={
        font=\small,
        at={(0.02,0.98)},
        anchor=north west,
        draw=none,
        fill =none
    },
    legend columns=1,
    legend cell align=left,
]

\addplot[black, solid, mark=triangle, mark repeat=5, mark phase=2]
table[
    x expr=\thisrow{y}*395,
    y expr=\thisrow{u_mean},
    col sep=comma,
    header=true
]{CF_InputData/C203_Linear_Re395_S0_RR.csv};
\addlegendentry{linear, $s=0$}

\addplot[ForestGreen, solid, mark=diamond, mark repeat=5, mark phase=3]
table[
    x expr=\thisrow{y}*395,
    y expr=\thisrow{u_mean},
    col sep=comma,
    header=true
]{CF_InputData/C204_Linear_Re395_S0P5_RR.csv};
\addlegendentry{linear, $s=0.5$}

\addplot[Purple, solid, mark=pentagon, mark repeat=5, mark phase=4]
table[
    x expr=\thisrow{y}*395,
    y expr=\thisrow{u_mean},
    col sep=comma,
    header=true
]{CF_InputData/C205_Linear_Re395_S1_RR.csv};
\addlegendentry{linear, $s=1$}

\addplot[blue, solid, mark=o, mark repeat=5, mark phase=0]
table[
    x expr=\thisrow{y}*395,
    y expr=\thisrow{u_mean},
    col sep=comma,
    header=true
]{CF_InputData/C201_nonLinear_Re395_S0_RR.csv};
\addlegendentry{nonlinear, $s=0$}

\addplot[red, solid, mark=square, mark repeat=5, mark phase=1]
table[
    x expr=\thisrow{y}*395,
    y expr=\thisrow{u_mean},
    col sep=comma,
    header=true
]{CF_InputData/C202_nonLinear_Re395_S1_RR.csv};
\addlegendentry{nonlinear, $s=1$}

\addplot[
    BurntOrange,
    very thick,
    dashdotted,
    mark=star,
    mark size=4pt,
    mark repeat=4,
    mark phase=3
]
table[
    x expr=\thisrow{y}*395,
    y expr=\thisrow{u_mean},
    col sep=comma,
    header=true
]{CF_InputData/Bazilevs_Linear_Re395_64.csv};
\addlegendentry{Bazilevs et al. (VMS)}

\addplot[
    black,
    very thick,
    mark=none
]
table[
    x=y+,
    y=Umean,
    col sep=comma,
    header=true
]{CF_InputData/chan395.means.csv};
\addlegendentry{Moser et al. (1999) DNS}

\end{axis}
\end{tikzpicture}
\caption{The mean velocity profile for $Re_\tau = 395$}
\label{fig:cf3d-re395-plots}
\end{figure}

\begin{figure}[!htb]
\centering
\input{CF_Plots/cf27_velocityFluctuations_Re395.tex}
\caption{Velocity fluctuations for $Re_\tau = 395$}
\label{fig:cf3d-re395-fluctuations}
\end{figure}

\subsection{Flow over an airfoil}\label{sec:airfoil}

We next simulate the three-dimensional turbulent flow over a NACA0012 airfoil~\cite{gregory1970low,mccroskey1987critical,ladson1988effects,hosseini2016direct} at the chord Reynolds number of $6\times 10^6$. The chord length of the airfoil is 1 and the angle of attack is 0 degrees. The boundary conditions used for the simulation are as follows (see \figref{fig:airfoilgeo}). The left ($x-$), top ($y+$) and bottom ($y-$) walls are prescribed with Dirichlet boundary conditions with \( u_x = 1 \), \( u_y= 0 \) and \( u_z = 0 \). The right boundary ($x+$) is prescribed with a traction-free boundary condition. The airfoil surface is prescribed with a no-slip boundary condition. The spanwise directions ($z-$ and $z+$) are set to be periodic.

\begin{figure}[!htb]
\centering
\begin{tikzpicture}[line cap=round, line join=round, scale=1.5,
                    x={(0.92cm,-0.22cm)}, y={(0cm,1cm)}, >=stealth]
\def\cx{0.5}     
\def\R{1.5}      
\def\Lx{4.0}     
\def\dx{0.5}    
\def\dy{0.85}
\begin{scope}[shift={(\dx,\dy)}]
\draw[thin]        (\cx,\R) arc[start angle=90, end angle=180, radius=\R];
\draw[thin,dashed] (\cx-\R,0) arc[start angle=180, end angle=270, radius=\R];
\draw[thin]        (\cx,\R) -- (\Lx,\R);
\draw[thin,dashed] (\cx,-\R) -- (\Lx,-\R);
\draw[thin]        (\Lx,-\R) -- (\Lx,\R);
\draw[thin,dashed]
plot[domain=0:1,samples=60,smooth]
(\x,{0.6*(0.2969*sqrt(\x)-0.1260*\x-0.3516*\x^2+0.2843*\x^3-0.1015*\x^4)})
-- plot[domain=1:0,samples=60,smooth]
(\x,{-0.6*(0.2969*sqrt(\x)-0.1260*\x-0.3516*\x^2+0.2843*\x^3-0.1015*\x^4)});
\end{scope}
\draw (\cx,\R)    -- ++(\dx,\dy);
\draw (\Lx,\R)    -- ++(\dx,\dy);
\draw (\Lx,-\R)   -- ++(\dx,\dy);
\draw[dashed] (\cx,-\R) -- ++(\dx,\dy);
\draw (\cx-\R,0)  -- ++(\dx,\dy);
\draw[thick] (\cx,\R) arc[start angle=90, end angle=270, radius=\R]
-- (\Lx,-\R) -- (\Lx,\R) -- cycle;
\draw[thick,fill=white]
plot[domain=0:1,samples=60,smooth]
(\x,{0.6*(0.2969*sqrt(\x)-0.1260*\x-0.3516*\x^2+0.2843*\x^3-0.1015*\x^4)})
-- plot[domain=1:0,samples=60,smooth]
(\x,{-0.6*(0.2969*sqrt(\x)-0.1260*\x-0.3516*\x^2+0.2843*\x^3-0.1015*\x^4)});

\draw (0,0) -- ++(\dx,\dy);
\draw (1,0) -- ++(\dx,\dy);

\node at (2.5, \R +0.2) [above] {Top ($y+$)};
\node at (2.5, -\R+0.6) [below] {Bottom ($y-$)};
\node at (\Lx+\dx+0.1, \dy/2) [right] {Right ($x+$)};
\node at (1.8, 0.1) {Front ($z+$)};
\node at (1.5, \R-0.3) {Back ($z-$)};
\foreach \y in {-1.0,-0.35,0.35,1.0}
\draw[->] (\cx-\R-1.05,\y) -- (\cx-\R-0.4,\y);
\node at (\cx-\R -0.5, 0.5) [above] {Left ($x-$)};
\node at (\cx-\R-1.1, 0) [left] {$u_x=1$};
\begin{scope}[shift={(-1.3,-2.6)}, scale=0.6]
\draw[very thick,->] (0,0) -- (1,0) node[right] {$x$};
\draw[very thick,->] (0,0) -- (0,1) node[above] {$y$};
\draw[very thick,->] (0,0) -- (-0.55,-0.60) node[below left] {$z$};
\end{scope}
\end{tikzpicture}
\caption{Schematic of the computational domain for the flow over a NACA0012 airfoil.}\label{fig:airfoilgeo}
\end{figure}

\begin{figure}[!htb]
\centering
\begin{subfigure}{0.49\textwidth}
    \centering
    \begin{minipage}{0.78\linewidth}
        \centering
        \includegraphics[width=\linewidth, trim={3in 1.5in 8in 1.5in},clip]{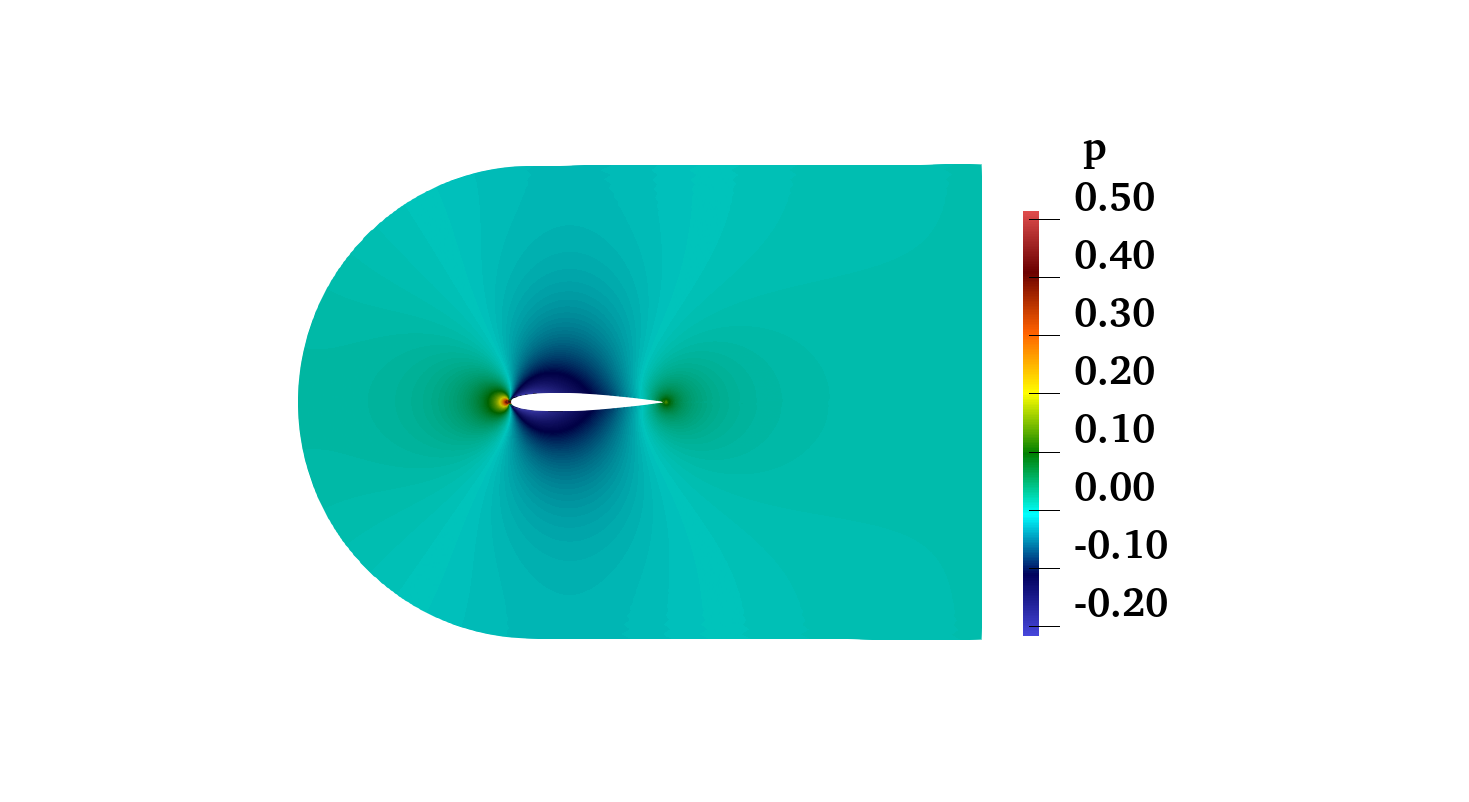}
    \end{minipage}%
    \hspace{-1.2cm}%
    \begin{minipage}{0.22\linewidth}
        \centering
        \begin{tikzpicture}
            \begin{axis}[
                hide axis,
                scale only axis,
                height=0pt,
                width=0pt,
                colormap={rainbowdesat}{
                    rgb=(0.278, 0.278, 0.859)
                    rgb=(0.000, 0.000, 0.360)
                    rgb=(0.000, 1.000, 1.000)
                    rgb=(0.000, 0.502, 0.000)
                    rgb=(1.000, 1.000, 0.000)
                    rgb=(1.000, 0.380, 0.380)
                    rgb=(0.420, 0.000, 0.000)
                },
                point meta min=-0.2,
                point meta max=0.5,
                colorbar,
                colorbar style={
                    xshift=1.5cm,
                    width=0.35cm,
                    height=4.5cm,
                    ytick={-0.2,-0.1,0.0,0.1,0.2,0.3,0.4,0.5},
                    yticklabel style={font=\small, /pgf/number format/.cd, fixed, fixed zerofill, precision=2},
                    title={$p$},
                    title style={font=\small},
                },
            ]
            \addplot [draw=none] coordinates {(0,0) (1,1)};
            \end{axis}
        \end{tikzpicture}
    \end{minipage}
    \caption{Pressure distribution}
    \label{fig:fpa-pressure}
\end{subfigure}%
\hfill
\begin{subfigure}{0.49\textwidth}
    \centering
    \begin{tikzpicture}
\begin{axis}[
    width=0.8\linewidth,
    xlabel={Chord length, $x/c$},
    ylabel={$C_p$},
    xmin=0.0,
    grid=both,
    grid style={dotted},
    line width=1pt,
    tick label style={font=\small},
    label style={font=\small},
    legend style={
        font=\small,
        at={(1,1)},
        anchor=north east,
        draw=none,
        fill=none
    },
    legend cell align=left,
]

\addplot[black, only marks, mark=none]
table[x=x, y=cp, col sep=comma, header=true]
{FPA_InputData/Experimental_Gregory_OReilly_cp.csv};
\addlegendentry{Gregory \& O'Reilly (1970)}

\addplot[blue, solid, mark=none, mark options={solid},mark repeat=2]
table[x=x, y=cp, col sep=comma, header=true]
{FPA_InputData/Case1_NonLinearS0_cp.csv}; 
\addlegendentry{nonlinear, $s=0$}

\addplot[red, solid, mark=none, mark options={solid},mark repeat=2]
table[x=x, y=cp, col sep=comma, header=true]
{FPA_InputData/Case2_NonLinearS1_cp.csv};
\addlegendentry{nonlinear, $s=1$}

\addplot[black, solid, mark=none, mark repeat=200]
table[x=x, y=cp, col sep=comma, header=true]
{FPA_InputData/Case3_LinearS0_cp.csv};
\addlegendentry{linear, $s=0$}

\addplot[ForestGreen, solid, mark=none, mark repeat=200]
table[x=x, y=cp, col sep=comma, header=true]
{FPA_InputData/Case4_LinearS0P5_cp.csv};
\addlegendentry{linear, $s=0.5$}

\addplot[Purple, solid, mark=none, mark repeat=200]
table[x=x, y=cp, col sep=comma, header=true]
{FPA_InputData/Case5_LinearS1_cp.csv};
\addlegendentry{linear, $s=1$}
\end{axis}
\end{tikzpicture}
    \caption{$C_p$ distribution}
    \label{fig:fpa-cp}
\end{subfigure}
\caption{Flow past an airfoil: (a) pressure field, (b) $C_p$ distribution along the airfoil surface.}
\label{fig:cp_plot}
\end{figure}

The computational domain shown in \figref{fig:airfoilgeo} extends 1 chord length upstream of the leading edge, 2 chord lengths downstream of the trailing edge and 1.5 chord lengths above and below the airfoil, with a spanwise extent of 0.1 chord lengths. The mesh consists of approximately $2\times 10^6$ nodes. The surface pressure coefficient ($C_p$) is compared with experimental measurements reported by Gregory and O'Reilly~\cite{gregory1970low}. The pressure coefficient ($C_p$) is calculated using
\begin{equation}
C_{p} = \frac{p - p_{\infty}}{\displaystyle \frac{1}{2} \rho_{\infty} u_{\infty}^{2}}.
\end{equation}
Here, $p_{\infty}$ is the freestream pressure, $\rho_{\infty}$ is the freestream density and $u_{\infty}$ is the freestream velocity. The free-stream conditions are \( u_{\infty} = 1 \), \( \rho_{\infty} = 1 \), and \( p_{\infty} = 0 \). Here $p$ is the mean pressure on the airfoil surface obtained from the simulation. The comparison is shown in \figref{fig:cp_plot} and we can see good agreement between all the linear and the nonlinear methods.

\subsection{Computation time}
\label{sec:timings}

\begin{figure}[!htb]
    \begin{minipage}{.49\textwidth}
		\centering
		\begin{tikzpicture}
			\begin{axis}[
				ybar,
				width=\linewidth,    
				symbolic x coords={
					L1,
					L2,
					L3,
					N1,
					N2
				},
				xtick=data,
                xtick align=inside,
				ymin=0,
				ylabel={$T^w_{\textrm{avg}}$(s)},
				bar width=5pt,
                grid=both,
    			grid style={dotted},
				legend style={at={(0.01,0.99)}, anchor=north west, font=\small},
				legend cell align=left,
				]
				\addplot+[fill=black!50] table[x=Re, y=solve_time, col sep=comma] {data/ldcGhia/solve_times_averages_combined_for_re_10000.txt};
                \node[anchor=north west, draw, fill=white, inner sep=3pt, font=\small, align=left] at (rel axis cs:0.02,0.98) {$N_v=16{,}641$\\$N_{\textrm{steps}}=5{,}000$};
			\end{axis}
		\end{tikzpicture}
		\subcaption{LDC (2D) $Re = 10^4$}
        \label{fig:ldc-re1000-solve-times}
	\end{minipage}
    \begin{minipage}{.49\textwidth}
		\centering
		\begin{tikzpicture}
			\begin{axis}[
				ybar,
				width=\linewidth,    
				symbolic x coords={
					L1,
					L2,
					L3,
					N1,
					N2
				},
				xtick=data,
                xtick align=inside,
				ymin=0,
				ylabel={$T^w_{\textrm{avg}}$(s)},
				bar width=5pt,
                grid=both,
    			grid style={dotted},
				legend style={at={(0.01,0.99)}, anchor=north west, font=\small},
				legend cell align=left,
				]
				\addplot+[fill=black!50] table[x=Re, y=solve_time, col sep=comma] {data/fpcKanaris/solve_times_averages_combined_for_re_300.txt};
                \node[anchor=north west, draw, fill=white, inner sep=3pt, font=\small, align=left] at (rel axis cs:0.02,0.98) {$N_v=45{,}279$\\$N_{\textrm{steps}}=8{,}000$};
			\end{axis}
		\end{tikzpicture}
		\subcaption{FPC (2D) $Re = 300$}
        \label{fig:fpc-re300-solve-times}
	\end{minipage}
    \\~\\~\\
    \begin{minipage}{.49\textwidth}
		\centering
		\begin{tikzpicture}
			\begin{axis}[
				ybar,
				width=\linewidth,
				height=0.9\linewidth,
				symbolic x coords={
					L1,
					L2,
					L3,
					N1,
					N2
				},
				xtick=data,
				xtick align=inside,
				ymin=0,
				ylabel={$T^w_{\textrm{avg}}$(s)},
				bar width=5pt,
				grid=both,
    			grid style={dotted},
				legend style={at={(0.01,0.99)}, anchor=north west, font=\small},
				legend cell align=left,
				]
				\addplot+[fill=black!50] table[x=nsf, y=solve_time, col sep=comma] {CF_InputData/solve_times_channelflow_re_395.txt};
                \node[anchor=north west, draw, fill=white, inner sep=3pt, font=\small, align=left] at (rel axis cs:0.02,0.98) {$N_v=274{,}625$\\$N_{\textrm{steps}}=28{,}000$};
			\end{axis}
		\end{tikzpicture}
		\subcaption{CF (3D) $Re_{\tau} = 395$}
        \label{fig:cf3-re395-solve-times}
	\end{minipage}
    \begin{minipage}{.49\textwidth}
		\centering
		\begin{tikzpicture}
			\begin{axis}[
				ybar,
				width=\linewidth,    
				height=0.9\linewidth, 
				symbolic x coords={
					L1,
					L2,
					L3,
					N1,
					N2
				},
				xtick=data,
				xtick align=inside,
				ymin=0,
				ylabel={$T^w_{\textrm{avg}}$(s)},
				bar width=5pt,
				grid=both,
    			grid style={dotted},
				]
				\addplot+[fill=black!50] table[x=nsf, y=solve_time, col sep=comma] {CF_InputData/solve_times_airfoil.txt};
                \node[anchor=north west, draw, fill=white, inner sep=3pt, font=\small, align=left] at (rel axis cs:0.02,0.98) {$N_v=2M$\\$N_{\textrm{steps}}=6{,}000$};
			\end{axis}
		\end{tikzpicture}
		\subcaption{FPA (3D) $Re = 6\times10^6$}
        \label{fig:fpa-re6e6-solve-times}
	\end{minipage}
    \caption{Average time taken (in wall clock time) per time step in the four benchmark cases (LDC: lid-driven cavity; FPC: flow past a circular cylinder; CF: turbulent channel flow; FPA: flow over an airfoil). Here, L1, L2 and L3 refer to the linear methods with $s=0,\ \halfnice$ and 1 respectively. Similarly, N1 and N2 stand for the nonlinear methods with $\dvc=0$ and 1 respectively. The number of mesh nodes $N_v$ and the number of steps $N_{\textrm{steps}}$ used for calculating the averages are indicated in the inset box in each panel.}
    \label{fig:ldc-fpc-solve-times}
\end{figure}
\begin{table}[!htb]
\centering
\caption{Number of Newton iterations (SNES iterations in \textsc{PETSc}) for the 3D cases with nonlinear formulation.}
\label{tab:cf3-fpa-snes-iters}
\begin{tabular}{c|c|c|c}
\toprule[2pt]
 &  & Channel flow & Flow past airfoil \\
\cmidrule(lr){3-3}\cmidrule(lr){4-4}
Formulation & $s$ & Re$_\tau$ = 395 & Re $= 6\times10^{6}$ \\
\midrule
Nonlinear & 0 & 3 & 3 \\
Nonlinear & 1 & 3 & 3 \\
\bottomrule[2pt]
\end{tabular}
\end{table}

\begin{table}[!htb]
\centering
\caption{Average number of Krylov subspace iterations required per time step for the 3D cases. The specified relative tolerance is set to $10^{-6}$ in all cases.}
\label{tab:cf3-fpa-ksp-iter-count}
\begin{tabular}{c|c|c|c}
\toprule[2pt]
 &  & Channel flow & Flow past airfoil \\
\cmidrule(lr){3-3}\cmidrule(lr){4-4}
Formulation & $s$ & Re$_\tau$ = 395 & Re $= 6\times10^{6}$ \\
\midrule
Linear    & 0            & 190.6 & 1126.9 \\
Linear    & $\halfnice$  & 191.3 & 1130.3 \\
Linear    & 1            & 191.5 & 1127.5 \\
\midrule
Nonlinear & 0            & 700.7 & 2095.7 \\
Nonlinear & 1            & 702.0 & 2062.4 \\
\bottomrule[2pt]
\end{tabular}
\end{table}

In the previous sections, we have established the accuracy and reliability of the proposed semi-implicit method across various standard benchmark problems. Now, in this section, we analyze their performance with respect to the time spent (in wall-clock units) in solving these problems. We collect the total solve time over a set number of time steps, and calculate the average solve time as $T^w_{\textrm{avg}} = T^w_{\textrm{total}} / N_{\textrm{steps}}$, where the $w$ denotes wall-clock time.

All the timings reported in this section are collected on the Frontera supercomputer at the Texas Advanced Computing Center (TACC)~\cite{stanzione2020frontera}. The simulations are run on the Cascade Lake (CLX) compute nodes; each CLX node has 56 cores with 192~GB of memory. The two-dimensional simulations (lid-driven cavity and flow past a cylinder) are run on a single CLX node using 36 MPI ranks, while the three-dimensional simulations (turbulent channel flow and flow over the airfoil) are run on 40 CLX nodes using 56 MPI ranks per node (2240 MPI ranks in total).

In \figref{fig:ldc-fpc-solve-times}, we present a comparison of $T^w_{\textrm{avg}}$ for the linear \eqref{eq:oseen-vms} and the nonlinear methods \eqref{eq:nse-vms}. The three linear methods L1, L2 and L3 denote $\dvc = 0,\halfnice,1$ respectively; and the nonlinear methods N1 and N2 denote $\dvc=0,1$ respectively. For each case, we have included the number of nodes in the mesh (denoted by $N_v$) and the total number of steps ($N_{\textrm{steps}}$) over which the wall-times are calculated. Within each case, the mesh and all other input data parameters are kept equal across different methods. We can see that the linear methods are roughly 2--4 times faster than the nonlinear methods. This gain is important when we consider the large number of time steps involved, especially in 3D problems. For example, if we focus on \figref{fig:cf3-re395-solve-times}, the total time taken by the nonlinear methods is roughly 46 hours as opposed to roughly $11$ hours by the linear cases. Similarly, in \figref{fig:fpa-re6e6-solve-times}, the total time taken by the nonlinear methods is roughly 47 hours whereas the linear methods take roughly 24 hours. 

\tabref{tab:cf3-fpa-snes-iters} shows that on average, roughly 3 Newton iterations are taken per time step in all the 3D cases (channel flow and airfoil). In \tabref{tab:cf3-fpa-ksp-iter-count}, we present the average number of KSP iterations (i.e., the Krylov subspace iterations) performed per time step. In the case of the linear methods, this relates to solving just one linear system, whereas, in the case of nonlinear methods, multiple successive linear systems are solved per time step. For the same mesh and parallelization, the time taken by each KSP iteration is roughly the same.

\section{Conclusions}
\label{sec:conclusions}

We presented a semi-implicit residual-based variational multiscale (VMS) formulation for the incompressible Navier--Stokes equations. The method linearizes the nonlinear convection by approximating the unknown convection with an extrapolated convection field. This Oseen-type approximation has a significant structural consequence in the VMS setting: the resulting linear advection operator admits an exact adjoint relation in the weak form, which enables the representation of convection-related fine-scale terms without introducing spatial derivatives of the modeled fine scales. This yields a uniform, implementation-ready stabilized formulation without case-dependent manipulations of the fine-scale derivatives.

The formulation is developed for a generalized linear advection operator
$[(\ddiv{\avec})\uvec + \dvc(\divergence\avec)\uvec]$ with $\dvc\in\{0,\tfrac12,1\}$, corresponding to the convective, skew-symmetric, and divergence forms. Across the numerical studies considered, the linear methods with $\dvc=0$ and $\dvc=\tfrac12$ were consistently accurate and robust. The $\dvc=1$ linear method performed comparably in settings with outlet and/or periodic boundary conditions, but it was less reliable in problems dominated by Dirichlet boundary conditions, where it did not exhibit the same convergence and/or needed additional refinement.

A practical advantage of the proposed semi-implicit approach is that each time step requires only a single linear solve. In our benchmarks, this translated to a $2$--$4\times$ reduction in wall-clock time relative to fully implicit nonlinear VMS while maintaining comparable accuracy in the regimes where the method is stable. This speedup is particularly relevant for parameter studies, design and optimization loops, and long-horizon computations that require a rapid approach to steady or statistically steady behavior.

Future work will focus on (i) considering the fine scales of the history velocities (that are currently discarded), (ii) extensions to complex-geometry discretizations and immersed-boundary approaches where the derivative-free fine-scale structure is especially advantageous, and (iii) assessing the method in large-scale three-dimensional turbulent simulations where solver scalability and time-to-solution are primary constraints.

\section*{Acknowledgments}
This work was partly supported by the National Science Foundation under award numbers 2053760 and 2436623. This support is gratefully acknowledged. We also acknowledge the Texas Advanced Computing Center (TACC) at The University of Texas at Austin for providing computational resources that have contributed to the research results reported within this paper.

\bibliographystyle{unsrt}
\bibliography{reflist}

\appendix
%
%

\section{Specialized forms}
\label{app:scheme-expansion}
Below, we specialize \eqref{eq:oseen-vms-bdf} for each $\dvc\in\{0,\tfrac12,1\}$. Since the three advective forms differ only in their contribution of the quantity $\divergence\avech$, it is instructive to identify the terms that contain $\divergence\avech$ in the final formulation. To this end, let us define
\begin{align}
	\divGreek := \divergence\avech.
\end{align}
Then the generalized advection operator is expressed as $\opAdvGen[\avech]{\dvc} = (\ddiv{\avech}) + \dvc\,\divGreek\,I$, and the operator appearing in its adjoint relation \eqref{eq:adv-weaken-relation} as $\opAdvGen[\avech]{1-\dvc} = (\ddiv{\avech}) + (1-\dvc)\,\divGreek\,I$, where $I$ denotes the identity operator. Throughout we use the $\tau$-weighted broken inner product
$\inner{\cdot}{\cdot}_{\tau,h} := \sum_{K\in\mesh}\inner{\cdot}{\tau\,\cdot}_K$
of \eqref{eq:weighted-broken-norm}, and terms carrying $\divGreek$ are
$\boxed{\text{boxed}}$.
\subsection{Convective form $(\dvc=0)$}

The trial and test operators become $\opAdvGen[\avech]{0} = (\ddiv{\avech})$ and $\opAdvGen[\avech]{1} = (\ddiv{\avech}) + \divGreek I$ respectively. Thus $\divGreek$ only appears in the stabilized terms. Given $\avech\in \spaceVhd$, we seek $(\uvechN, \phN)\in \spaceVhd\times\spaceQh$ such that for all $(\vvech, \qh) \in \spaceVh\times \spaceQh$,
\begin{align}
  \label{eq:app-scheme-zero}
  &\inner{\vvech}{\sigma\uvechN}
   + \inner{\vvech}{(\ddiv{\avech})\uvechN}
   + \visco\inner{\grad\vvech}{\grad\uvechN}
   - \inner{\divergence\vvech}{\phN}
  \notag\\
  &\quad
   + \inner{\qh}{\divergence\uvechN}
   + \inner{\divergence\vvech}{\divergence\uvechN}_{\tauc,h}
  \notag\\
  &\quad
   + \inner{(\ddiv{\avech})\vvech}{\sigma\uvechN}_{\taum,h}
   + \inner{(\ddiv{\avech})\vvech}{(\ddiv{\avech})\uvechN}_{\taum,h}
  \notag\\
  &\quad
   - \visco\inner{(\ddiv{\avech})\vvech}{\laplacian\uvechN}_{\taum,h}
   + \inner{(\ddiv{\avech})\vvech}{\grad\phN}_{\taum,h}
  \notag\\
  &\quad
   + \boxed{\inner{\divGreek\,\vvech}{\sigma\uvechN}_{\taum,h}}
   + \boxed{\inner{\divGreek\,\vvech}{(\ddiv{\avech})\uvechN}_{\taum,h}}
  \notag\\
  &\quad
   - \visco\boxed{\inner{\divGreek\,\vvech}{\laplacian\uvechN}_{\taum,h}}
   + \boxed{\inner{\divGreek\,\vvech}{\grad\phN}_{\taum,h}}
  \notag\\
  &\quad
   + \inner{\grad\qh}{\sigma\uvechN}_{\taum,h}
   + \inner{\grad\qh}{(\ddiv{\avech})\uvechN}_{\taum,h}
  \notag\\
  &\quad
   - \visco\inner{\grad\qh}{\laplacian\uvechN}_{\taum,h}
   + \inner{\grad\qh}{\grad\phN}_{\taum,h}
  \notag\\
  &=
   \inner{\vvech}{\gvecN}
   + \inner{(\ddiv{\avech})\vvech}{\gvecN}_{\taum,h}
   + \boxed{\inner{\divGreek\,\vvech}{\gvecN}_{\taum,h}}
   + \inner{\grad\qh}{\gvecN}_{\taum,h}
   + \inner{\vvech}{\hvec}_{\Gamma_{N}}.
\end{align}

\subsection{Skew-symmetric form $(\dvc=\halfnice)$}

Both test and trial operators become $\opAdvGen[\avech]{\halfnice} = (\ddiv{\avech}) + \halfnice\, \divGreek I$ respectively. Given $\avech\in \spaceVhd$, we seek $(\uvechN, \phN)\in \spaceVhd\times\spaceQh$ such that for all $(\vvech, \qh) \in \spaceVh\times \spaceQh$,
\begin{align}
  \label{eq:app-scheme-half}
  &\inner{\vvech}{\sigma\uvechN}
   + \inner{\vvech}{(\ddiv{\avech})\uvechN}
   + \tfrac12\boxed{\inner{\vvech}{\divGreek\,\uvechN}}
   + \visco\inner{\grad\vvech}{\grad\uvechN}
  \notag\\
  &\quad
   - \inner{\divergence\vvech}{\phN}
   + \inner{\qh}{\divergence\uvechN}
   + \inner{\divergence\vvech}{\divergence\uvechN}_{\tauc,h}
  \notag\\
  &\quad
   + \inner{(\ddiv{\avech})\vvech}{\sigma\uvechN}_{\taum,h}
   + \inner{(\ddiv{\avech})\vvech}{(\ddiv{\avech})\uvechN}_{\taum,h}
   + \tfrac12\boxed{\inner{(\ddiv{\avech})\vvech}{\divGreek\,\uvechN}_{\taum,h}}
  \notag\\
  &\quad
   - \visco\inner{(\ddiv{\avech})\vvech}{\laplacian\uvechN}_{\taum,h}
   + \inner{(\ddiv{\avech})\vvech}{\grad\phN}_{\taum,h}
  \notag\\
  &\quad
   + \tfrac12\boxed{\inner{\divGreek\,\vvech}{\sigma\uvechN}_{\taum,h}}
   + \tfrac12\boxed{\inner{\divGreek\,\vvech}{(\ddiv{\avech})\uvechN}_{\taum,h}}
   + \tfrac14\boxed{\inner{\divGreek\,\vvech}{\divGreek\,\uvechN}_{\taum,h}}
  \notag\\
  &\quad
   - \tfrac12\visco\boxed{\inner{\divGreek\,\vvech}{\laplacian\uvechN}_{\taum,h}}
   + \tfrac12\boxed{\inner{\divGreek\,\vvech}{\grad\phN}_{\taum,h}}
  \notag\\
  &\quad
   + \inner{\grad\qh}{\sigma\uvechN}_{\taum,h}
   + \inner{\grad\qh}{(\ddiv{\avech})\uvechN}_{\taum,h}
   + \tfrac12\boxed{\inner{\grad\qh}{\divGreek\,\uvechN}_{\taum,h}}
  \notag\\
  &\quad
   - \visco\inner{\grad\qh}{\laplacian\uvechN}_{\taum,h}
   + \inner{\grad\qh}{\grad\phN}_{\taum,h}
  \notag\\
  &=
   \inner{\vvech}{\gvecN}
   + \inner{(\ddiv{\avech})\vvech}{\gvecN}_{\taum,h}
   + \tfrac12\boxed{\inner{\divGreek\,\vvech}{\gvecN}_{\taum,h}}
   + \inner{\grad\qh}{\gvecN}_{\taum,h}
   + \inner{\vvech}{\hvec}_{\Gamma_{N}} .
\end{align}

\subsection{Divergence form $(\dvc=1)$}

The test operator reduces to $\opAdvGen[\avech]{0}=(\ddiv{\avech})$. The trial operator becomes $\opAdvGen[\avech]{1} = (\ddiv{\avech}) + \divGreek I$. Thus $\divGreek$ only appears in the stabilized terms. Given $\avech\in \spaceVhd$, we seek $(\uvechN, \phN)\in \spaceVhd\times\spaceQh$ such that for all $(\vvech, \qh) \in \spaceVh\times \spaceQh$,
\begin{align}
  \label{eq:app-scheme-one}
  &\inner{\vvech}{\sigma\uvechN}
   + \inner{\vvech}{(\ddiv{\avech})\uvechN}
   + \boxed{\inner{\vvech}{\divGreek\,\uvechN}}
   + \visco\inner{\grad\vvech}{\grad\uvechN}
  \notag\\
  &\quad
   - \inner{\divergence\vvech}{\phN}
   + \inner{\qh}{\divergence\uvechN}
   + \inner{\divergence\vvech}{\divergence\uvechN}_{\tauc,h}
  \notag\\
  &\quad
   + \inner{(\ddiv{\avech})\vvech}{\sigma\uvechN}_{\taum,h}
   + \inner{(\ddiv{\avech})\vvech}{(\ddiv{\avech})\uvechN}_{\taum,h}
   + \boxed{\inner{(\ddiv{\avech})\vvech}{\divGreek\,\uvechN}_{\taum,h}}
  \notag\\
  &\quad
   - \visco\inner{(\ddiv{\avech})\vvech}{\laplacian\uvechN}_{\taum,h}
   + \inner{(\ddiv{\avech})\vvech}{\grad\phN}_{\taum,h}
  \notag\\
  &\quad
   + \inner{\grad\qh}{\sigma\uvechN}_{\taum,h}
   + \inner{\grad\qh}{(\ddiv{\avech})\uvechN}_{\taum,h}
   + \boxed{\inner{\grad\qh}{\divGreek\,\uvechN}_{\taum,h}}
  \notag\\
  &\quad
   - \visco\inner{\grad\qh}{\laplacian\uvechN}_{\taum,h}
   + \inner{\grad\qh}{\grad\phN}_{\taum,h}
  \notag\\
  &=
   \inner{\vvech}{\gvecN}
   + \inner{(\ddiv{\avech})\vvech}{\gvecN}_{\taum,h}
   + \inner{\grad\qh}{\gvecN}_{\taum,h}
   + \inner{\vvech}{\hvec}_{\Gamma_{N}} .
\end{align}

\subsection{The differing terms in isolation}
\label{app:scheme-diff}
Here, we isolate the $\divGreek$-dependent terms of the formulation in \eqref{eq:oseen-vms-bdf}. Moving all the terms in \eqref{eq:oseen-vms-bdf} to the left hand side and collating the terms containing $\divGreek$, we can define
\begin{align}
	\label{eq:app-theta-general}
	\Psi^{(\dvc)}
	&:= (1-\dvc)\Bigl[\,
	\inner{\divGreek\,\vvech}{\sigma\uvechN}_{\taum,h}
	+ \inner{\divGreek\,\vvech}{(\ddiv{\avech})\uvechN}_{\taum,h}
	\notag\\
	&\hspace{4.2em}
	- \visco\inner{\divGreek\,\vvech}{\laplacian\uvechN}_{\taum,h}
	+ \inner{\divGreek\,\vvech}{\grad\phN}_{\taum,h}
	- \inner{\divGreek\,\vvech}{\gvecN}_{\taum,h}
	\Bigr]
	\notag\\[4pt]
	&\quad
	+ \dvc\Bigl[\,
	\inner{\vvech}{\divGreek\,\uvechN}
	+ \inner{(\ddiv{\avech})\vvech}{\divGreek\,\uvechN}_{\taum,h}
	+ \inner{\grad\qh}{\divGreek\,\uvechN}_{\taum,h}
	\Bigr]
	\notag\\[4pt]
	&\quad
	+ \dvc(1-\dvc)\,\inner{\divGreek\,\vvech}{\divGreek\,\uvechN}_{\taum,h}.
\end{align}
Note that the five terms in the first bracket are precisely the components of $\resMom^{(0)}$ tested against $\divGreek\,\vvech$, so that bracket is $\inner{\divGreek\,\vvech}{\resMom^{(0)}}_{\taum,h}$. Substituting $\dvc = 0,1,\halfnice$ in that order now gives the individual forms:
\begin{subequations}
\label{eq:app-theta}
\begin{align}
  \label{eq:app-theta-zero}
  \Psi^{(0)}
  &= \inner{\divGreek\,\vvech}{\resMom^{(0)}}_{\taum,h} ,
  \\[4pt]
  \label{eq:app-theta-one}
  \Psi^{(1)}
  &= \; \inner{\vvech}{\divGreek\,\uvechN}
  + \inner{(\ddiv{\avech})\vvech}{\divGreek\,\uvechN}_{\taum,h}
  + \inner{\grad\qh}{\divGreek\,\uvechN}_{\taum,h} ,
  \\[4pt]
  \label{eq:app-theta-half}
  \Psi^{(\halfnice)}
  &= \tfrac12 \Psi^{(0)} + \tfrac12 \Psi^{(1)}
   + \tfrac14\inner{\divGreek\,\vvech}{\divGreek\,\uvechN}_{\taum,h}
  .
\end{align}
\end{subequations}
Note that $\divGreek$ enters only the momentum equations for $\dvc=0$, whereas for $\dvc=\halfnice, 1$, it also enters the continuity equation.

%

\section{Derivation of the discrete energy balance}
\label{app:energy-derivation}

In this appendix we prove Proposition~\ref{thm:energy-discrete} and the
specializations to $\dvc=0,\halfnice,1$ as stated in
Section~\ref{sec:energy-discrete}. Throughout, $\resMom^{(\dvc)}$ is the
momentum residual \eqref{eq:res-def}, $\bdfop$ is the BDF operator
\eqref{eq:bdf-operator}, homogeneous Dirichlet data are assumed, and the bounds
\eqref{eq:tau-bounds}, the inverse estimate \eqref{eq:inverse-estimate} and the
condition \eqref{eq:cond-inv} are in force. Since
$\opAdvGen[\avech]{1}\uvec = \opAdvGen[\avech]{0}\uvec + (\divergence\avech)\uvec$,
the residuals for $\dvc=0$ and $\dvc=1$ satisfy
\begin{align}
  \label{eq:res-shift}
  \resMom^{(1)} = \resMom^{(0)} + (\divergence\avech)\,\uvechN
\end{align}
(the data $\fvecN$ cancels in the difference).

\subsection{Testing by the trial functions}
\label{app:energy-testing}
Substitute $(\vvech,\qh)=(\uvechN,\phN)$ in \eqref{eq:oseen-vms-bdf} and move
every occurrence of $\gvecN$ to the left-hand side.

\smallskip
\noindent\textit{Temporal and data terms.}\quad
By the regrouping identity in \eqref{eq:bdf-operator},
$\inner{\uvechN}{\sigma\uvechN} - \inner{\uvechN}{\gvecN}
 = \inner{\bdfop\uvechN}{\uvechN} - \inner{\fvecN}{\uvechN}$.

\smallskip
\noindent\textit{Viscous term:}\quad
$\visco\|\grad\uvechN\|^2 \geq 0$.

\smallskip
\noindent\textit{Pressure terms:}\quad
$-\inner{\divergence\uvechN}{\phN} + \inner{\phN}{\divergence\uvechN} = 0$.

\smallskip
\noindent\textit{Grad-div term:}\quad
$\norm{\divergence\uvechN}_{\tauc,h}^2 \geq 0$.

\smallskip
\noindent\textit{Coarse-scale convection.}\quad
Since $\uvechN$ vanishes on $\partial\spDom$, Proposition~\ref{thm:energy-identity} gives
\begin{align}
	\inner{\uvechN}{\opAdvGen[\avech]{\dvc}\uvechN}
	= \left(\dvc-\tfrac{1}{2}\right)\inner{(\divergence\avech)\,\uvechN}{\uvechN}
	= T_1^{(\dvc)} .
\end{align}

\smallskip
\noindent\textit{Stabilized terms.}\quad
The data $\gvecN$ enters \eqref{eq:oseen-vms-bdf} twice on the right-hand side:
through $\inner{\vvech}{\gvecN}$, treated above, and through
$\sum_{K}\inner{\taum\opAdvGen[\avech]{1-\dvc}\vvech+\taum\grad\qh}{\gvecN}_K$.
Moving the latter to the left and combining it with the residual-based
stabilization terms, the whole stabilized block tests against the full
residual \eqref{eq:res-def}:
\begin{align}
  \label{eq:supg-block}
  \sum_{K\in\mesh}
  \inner{\opAdvGen[\avech]{1-\dvc}\uvechN+\grad\phN}{\taum\resMom^{(\dvc)}}_K .
\end{align}
Using $\opAdvGen[\avech]{1-\dvc}-\opAdvGen[\avech]{\dvc}=(1-2\dvc)(\divergence\avech)I$
and \eqref{eq:res-def},
\begin{align}
  \label{eq:supg-rhs-expand}
  \opAdvGen[\avech]{1-\dvc}\uvechN + \grad\phN
  \;=\;
  \resMom^{(\dvc)}
  - \bdfop\uvechN
  + \visco\laplacian\uvechN
  + (1-2\dvc)(\divergence\avech)\,\uvechN
  + \fvecN .
\end{align}
Dotting with $\taum\resMom^{(\dvc)}$ and summing over elements, \eqref{eq:supg-block} equals
\begin{align}
  \label{eq:supg-four}
  \norm{\resMom^{(\dvc)}}_{\taum,h}^2
  \;-\; T_3 \;+\; T_4 \;+\; T_2^{(\dvc)} \;+\; T_5 ,
\end{align}
with $T_2^{(\dvc)}$ as in \eqref{eq:T1-T2-def} and
\begin{align}
  \label{eq:T3-T5-def}
  T_3 := \inner{\bdfop\uvechN}{\resMom^{(\dvc)}}_{\taum,h},
  \quad
  T_4 := \visco\inner{\laplacian\uvechN}{\resMom^{(\dvc)}}_{\taum,h},
  \quad
  T_5 := \inner{\fvecN}{\resMom^{(\dvc)}}_{\taum,h} .
\end{align}
Collecting the terms, every solution of \eqref{eq:oseen-vms-bdf} satisfies the identity
\begin{multline}
  \label{eq:balance-identity}
  \inner{\bdfop\uvechN}{\uvechN}
  + \visco\|\grad\uvechN\|^2
  + \norm{\divergence\uvechN}_{\tauc,h}^2
  + \norm{\resMom^{(\dvc)}}_{\taum,h}^2
  + T_1^{(\dvc)} + T_2^{(\dvc)} \\
  \;=\; \inner{\fvecN}{\uvechN} + T_3 - T_4 - T_5 .
\end{multline}

\subsection{Bounding the cross-terms}
\label{app:energy-crossterms}

\noindent\textit{Cross-term $T_3$.}\quad
Cauchy--Schwarz and Young's inequality with parameter $\tfrac14$, followed by
the first bound in \eqref{eq:tau-bounds}:
\begin{align}
  \label{eq:T3-bd}
  |T_3|
  \;\leq\; \tfrac14\norm{\resMom^{(\dvc)}}_{\taum,h}^2
  + \norm{\bdfop\uvechN}_{\taum,h}^2
  \;\leq\; \tfrac14\norm{\resMom^{(\dvc)}}_{\taum,h}^2
  + \frac{\dt}{2\beta_0}\norm{\bdfop\uvechN}^2 .
\end{align}

\smallskip
\noindent\textit{Cross-term $T_4$.}\quad
Element-wise Cauchy--Schwarz, Young's inequality with parameter $\halfnice$,
the inverse estimate \eqref{eq:inverse-estimate}, and then the second bound in
\eqref{eq:tau-bounds}:
\begin{align*}
|T_4|
 &\leq \tfrac12\visco^2\norm{\laplacian\uvechN}_{\taum,h}^2
      + \tfrac12\norm{\resMom^{(\dvc)}}_{\taum,h}^2
  \;\leq\; \tfrac12\sum_{K\in\mesh}\visco^2\,\cinv\, h_K^{-2}
       \norm{\taum}_{L^\infty(K)}\norm{\grad\uvechN}_K^2
      + \tfrac12\norm{\resMom^{(\dvc)}}_{\taum,h}^2 \\
 &\leq \frac{\visco}{8}\sqrt{\frac{\cinv}{d}}\;\norm{\grad\uvechN}^2
      + \tfrac12\norm{\resMom^{(\dvc)}}_{\taum,h}^2 .
\end{align*}
With \eqref{eq:cond-inv} this gives
\begin{align}
  \label{eq:T4-bd}
  |T_4| \leq \frac{\alpha\,\visco}{8}\,\|\grad\uvechN\|^2
             + \tfrac12\norm{\resMom^{(\dvc)}}_{\taum,h}^2 ,
\end{align}
and the requirement $\alpha<8$ is precisely what leaves a positive fraction of
the viscous term $\visco\norm{\grad\uvechN}^2$ once $T_4$ has been absorbed.

\smallskip
\noindent\textit{Cross-term $T_5$.}\quad
Cauchy--Schwarz, Young with parameter $\tfrac18$, then \eqref{eq:tau-bounds}:
\begin{align}
  \label{eq:T5-bd}
  |T_5| \;\leq\; 2\norm{\fvecN}_{\taum,h}^2
  + \tfrac18\norm{\resMom^{(\dvc)}}_{\taum,h}^2
  \;\leq\; \frac{\dt}{\beta_0}\|\fvecN\|^2
  + \tfrac18\norm{\resMom^{(\dvc)}}_{\taum,h}^2 .
\end{align}

\smallskip
\noindent\textit{Proof of Proposition~\ref{thm:energy-discrete}.}\quad
Insert \eqref{eq:T3-bd}--\eqref{eq:T5-bd} into \eqref{eq:balance-identity}. The
three cross-terms consume $\tfrac14+\tfrac12+\tfrac18=\tfrac78$ of the residual
self-term, leaving $\tfrac18\norm{\resMom^{(\dvc)}}_{\taum,h}^2$, and a fraction
$\alpha/8$ of the viscous term, leaving
$\left(1-\tfrac{\alpha}{8}\right)\visco\norm{\grad\uvechN}^2$.
What remains is exactly \eqref{eq:energy-general}. \hfill$\square$

\subsection{Specialization to the three forms}
\label{app:energy-specialization}
Write $\varsigma := \divergence\avech$ and
$b := \norm{\varsigma\,\uvechN}_{\taum,h}$, and collect the parts of
\eqref{eq:energy-general} carrying no explicit factor of $\dvc$ as
\begin{align}
  \mathcal{L} := \inner{\bdfop\uvechN}{\uvechN}
  + \left(1-\frac{\alpha}{8}\right)\visco\,\|\grad\uvechN\|^2
  + \norm{\divergence\uvechN}_{\tauc,h}^2 ,
  \qquad
  \mathcal{R} := \frac{\dt}{2\beta_0}\norm{\bdfop\uvechN}^2
  + \inner{\fvecN}{\uvechN} + \frac{\dt}{\beta_0}\|\fvecN\|^2 .
\end{align}

\smallskip
\noindent\textit{Skew-symmetric form} ($\dvc=\halfnice$).\quad
Both $T_1^{(1/2)}$ and $T_2^{(1/2)}$ carry the factor $\dvc-\halfnice$ or
$1-2\dvc$ and therefore vanish identically, so \eqref{eq:energy-general} reads
\begin{align}
  \label{eq:energy-half}
  \mathcal{L} + \tfrac18\norm{\resMom^{(1/2)}}_{\taum,h}^2 \;\leq\; \mathcal{R} ,
\end{align}
in which every term on the left is non-negative.

\smallskip
\noindent\textit{Convective form} ($\dvc=0$).\quad
Here $T_1^{(0)} = -\tfrac12\inner{\varsigma\uvechN}{\uvechN}$ and
$T_2^{(0)} = \inner{\varsigma\uvechN}{\resMom^{(0)}}_{\taum,h}$, so
\begin{align}
  \label{eq:energy-zero}
  \mathcal{L}
  + \underbrace{\tfrac18\norm{\resMom^{(0)}}_{\taum,h}^2
    + \inner{\varsigma\uvechN}{\resMom^{(0)}}_{\taum,h}}_{\geq\;-2b^2}
  - \tfrac12\inner{\varsigma\uvechN}{\uvechN}
  \;\leq\; \mathcal{R} ,
\end{align}
the underbrace following from
$\min_{a\geq0}\big(\tfrac18a^2-ab\big) = -2b^2$, attained at $a=4b$.

\smallskip
\noindent\textit{Divergence form} ($\dvc=1$).\quad
Here $T_1^{(1)} = +\tfrac12\inner{\varsigma\uvechN}{\uvechN}$ and
$T_2^{(1)} = -\inner{\varsigma\uvechN}{\resMom^{(1)}}_{\taum,h}$, so
\begin{align}
  \label{eq:energy-one}
  \mathcal{L}
  + \underbrace{\tfrac18\norm{\resMom^{(1)}}_{\taum,h}^2
    - \inner{\varsigma\uvechN}{\resMom^{(1)}}_{\taum,h}}_{\geq\;-2b^2}
  + \tfrac12\inner{\varsigma\uvechN}{\uvechN}
  \;\leq\; \mathcal{R} .
\end{align}

\smallskip
\noindent\textit{Comparison.}\quad
The two underbraced quantities in \eqref{eq:energy-zero} and
\eqref{eq:energy-one} have the same sharp lower bound $-2b^2$, the only
difference being whether it is attained at $\resMom^{(0)}=4\varsigma\uvechN$ or
at $\resMom^{(1)}=-4\varsigma\uvechN$. The estimate therefore separates
$\dvc=\halfnice$, for which the $\varsigma$-coupling is absent altogether, from
$\dvc\in\{0,1\}$.

\end{document}